\documentclass[12pt]{article}
\usepackage{epsfig}
\usepackage{amsmath}
\usepackage{color}
\usepackage{colortbl}
\usepackage{hhline}
\usepackage{amssymb}
\usepackage{times}
\usepackage{cite}
\usepackage{array}

\newlength{\dinwidth}
\newlength{\dinmargin}
\begin{document}  
\newcommand{\pom}{{I\!\!P}}
\newcommand{\reg}{{I\!\!R}}
\newcommand{\slowpi}{\pi_{\mathit{slow}}}
\newcommand{\fiidiii}{F_2^{D(3)}}
\newcommand{\fiidiiiarg}{\fiidiii\,(\beta,\,Q^2,\,x)}
\newcommand{\n}{1.19\pm 0.06 (stat.) \pm0.07 (syst.)}
\newcommand{\nz}{1.30\pm 0.08 (stat.)^{+0.08}_{-0.14} (syst.)}
\newcommand{\fiidiiiful}{F_2^{D(4)}\,(\beta,\,Q^2,\,x,\,t)}
\newcommand{\fiipom}{\tilde F_2^D}
\newcommand{\ALPHA}{1.10\pm0.03 (stat.) \pm0.04 (syst.)}
\newcommand{\ALPHAZ}{1.15\pm0.04 (stat.)^{+0.04}_{-0.07} (syst.)}
\newcommand{\fiipomarg}{\fiipom\,(\beta,\,Q^2)}
\newcommand{\pomflux}{f_{\pom / p}}
\newcommand{\nxpom}{1.19\pm 0.06 (stat.) \pm0.07 (syst.)}
\newcommand {\gapprox}
   {\raisebox{-0.7ex}{$\stackrel {\textstyle>}{\sim}$}}
\newcommand {\lapprox}
   {\raisebox{-0.7ex}{$\stackrel {\textstyle<}{\sim}$}}
\def\gsim{\,\lower.25ex\hbox{$\scriptstyle\sim$}\kern-1.30ex%
\raise 0.55ex\hbox{$\scriptstyle >$}\,}
\def\lsim{\,\lower.25ex\hbox{$\scriptstyle\sim$}\kern-1.30ex%
\raise 0.55ex\hbox{$\scriptstyle <$}\,}
\newcommand{\pomfluxarg}{f_{\pom / p}\,(x_\pom)}
\newcommand{\dsf}{\mbox{$F_2^{D(3)}$}}
\newcommand{\dsfva}{\mbox{$F_2^{D(3)}(\beta,Q^2,x_{I\!\!P})$}}
\newcommand{\dsfvb}{\mbox{$F_2^{D(3)}(\beta,Q^2,x)$}}
\newcommand{\dsfpom}{$F_2^{I\!\!P}$}
\newcommand{\gap}{\stackrel{>}{\sim}}
\newcommand{\lap}{\stackrel{<}{\sim}}
\newcommand{\fem}{$F_2^{em}$}
\newcommand{\tsnmp}{$\tilde{\sigma}_{NC}(e^{\mp})$}
\newcommand{\tsnm}{$\tilde{\sigma}_{NC}(e^-)$}
\newcommand{\tsnp}{$\tilde{\sigma}_{NC}(e^+)$}
\newcommand{\st}{$\star$}
\newcommand{\sst}{$\star \star$}
\newcommand{\ssst}{$\star \star \star$}
\newcommand{\sssst}{$\star \star \star \star$}
\newcommand{\tw}{\theta_W}
\newcommand{\sw}{\sin{\theta_W}}
\newcommand{\cw}{\cos{\theta_W}}
\newcommand{\sww}{\sin^2{\theta_W}}
\newcommand{\cww}{\cos^2{\theta_W}}
\newcommand{\trm}{m_{\perp}}
\newcommand{\trp}{p_{\perp}}
\newcommand{\trmm}{m_{\perp}^2}
\newcommand{\trpp}{p_{\perp}^2}
\newcommand{\alp}{\alpha_s}

\newcommand{\alps}{\alpha_s}
\newcommand{\sqrts}{$\sqrt{s}$}
\newcommand{\LO}{$O(\alpha_s^0)$}
\newcommand{\Oa}{$O(\alpha_s)$}
\newcommand{\Oaa}{$O(\alpha_s^2)$}
\newcommand{\PT}{p_{\perp}}
\newcommand{\JPSI}{J/\psi}
\newcommand{\sh}{\hat{s}}
\newcommand{\uh}{\hat{u}}
\newcommand{\MP}{m_{J/\psi}}
\newcommand{\PO}{I\!\!P}
\newcommand{\xbj}{x}
\newcommand{\xpom}{x_{\PO}}
\newcommand{\ttbs}{\char'134}
\newcommand{\xpomlo}{3\times10^{-4}}  
\newcommand{\xpomup}{0.05}  
\newcommand{\dgr}{^\circ}
\newcommand{\pbarnt}{\,\mbox{{\rm pb$^{-1}$}}}
\newcommand{\gev}{\,\mbox{GeV}}
\newcommand{\WBoson}{\mbox{$W$}}
\newcommand{\fbarn}{\,\mbox{{\rm fb}}}
\newcommand{\fbarnt}{\,\mbox{{\rm fb$^{-1}$}}}
%
%
\newcommand{\qsq}{\ensuremath{Q^2} }
\newcommand{\gevsq}{\ensuremath{\mathrm{GeV}^2} }
\newcommand{\et}{\ensuremath{E_t^*} }
\newcommand{\rap}{\ensuremath{\eta^*} }
\newcommand{\gp}{\ensuremath{\gamma^*}p }
\newcommand{\dsiget}{\ensuremath{{\rm d}\sigma_{ep}/{\rm d}E_t^*} }
\newcommand{\dsigrap}{\ensuremath{{\rm d}\sigma_{ep}/{\rm d}\eta^*} }

\newcommand{\hdick}{\noalign{\hrule height1.4pt}}
\newcommand{\up}{\raisebox{1.7ex}[-1.7ex]}

\definecolor{indianred}{rgb}{0.5812,0.0665,0.0659} 
\definecolor{orange}{rgb}{1.0,0.796875,0.2578125} 

\def\Journal#1#2#3#4{{#1} {\bf #2} (#3) #4}
\def\NCA{\em Nuovo Cimento}
\def\NIM{\em Nucl. Instrum. Methods}
\def\NIMA{{\em Nucl. Instrum. Methods} {\bf A}}
\def\NPB{{\em Nucl. Phys.}   {\bf B}}
\def\PLB{{\em Phys. Lett.}   {\bf B}}
\def\PRL{\em Phys. Rev. Lett.}
\def\PRD{{\em Phys. Rev.}    {\bf D}}
\def\ZPC{{\em Z. Phys.}      {\bf C}}
\def\EJC{{\em Eur. Phys. J.} {\bf C}}
\def\CPC{\em Comp. Phys. Commun.}

\begin{titlepage}

\noindent

\begin{flushleft}
{\tt DESY 07-009    \hfill    ISSN 0418-9833} \\
{\tt February  2007}                  \\
\end{flushleft}

\vspace*{3cm}

\begin{center}
\begin{Large}

{\bf Search for Lepton Flavour Violation \\
    in \boldmath{$ep$} Collisions at HERA\\}

\vspace{2cm}

H1 Collaboration

\end{Large}
\end{center}

\vspace{2cm}

\begin{abstract} \noindent
A search for the lepton flavour violating processes \mbox{$ep\rightarrow \mu
  X$} and \mbox{$ep\rightarrow \tau X$} is 
performed with the H1 experiment 
at HERA. Final states with a muon or tau and a hadronic jet are 
searched for in a data sample corresponding to an integrated luminosity of 
$66.5\,{\rm pb^{-1}}$ for $e^{+}p$ collisions and 
$13.7\,{\rm pb^{-1}}$ for $e^{-}p$ collisions at a centre-of-mass energy of 
$319\,{\rm GeV}$. No evidence 
for lepton flavour violation is found. 
Limits are derived on the mass and the couplings of leptoquarks inducing lepton flavour violation 
in an extension of the Buchm\"uller-R\"uckl-Wyler effective model. 
Leptoquarks produced in $ep$ collisions with a coupling strength of 
$\lambda=0.3$ and decaying with the same coupling strength 
to a muon-quark pair or a tau-quark pair are excluded at 
95\% confidence level up to masses of $459\,{\rm GeV}$ and 
$379\,{\rm GeV}$, respectively. 
\end{abstract}

\vspace{1.5cm}

\begin{center}
Submitted to \EJC
\end{center}

\end{titlepage}

%
%
%
\newpage 
\noindent
A.~Aktas$^{10}$,               
C.~Alexa$^{10,49}$,            
V.~Andreev$^{24}$,             
T.~Anthonis$^{4}$,             
B.~Antunovic$^{25}$,           
S.~Aplin$^{10}$,               
A.~Asmone$^{32}$,              
A.~Astvatsatourov$^{4}$,       
A.~Babaev$^{23, \dagger}$,     
S.~Backovic$^{29}$
% ...

\bigskip{\it \noindent
% ...
$ ^{4}$ Inter-University Institute for High Energies ULB-VUB, Brussels;
          Universiteit Antwerpen, Antwerpen; Belgium$^{ c}$ \\
% ...
$ ^{10}$ DESY, Hamburg, Germany \\
% ...
$ ^{23}$ Institute for Theoretical and Experimental Physics,
          Moscow, Russia$^{ k}$ \\
 $ ^{24}$ Lebedev Physical Institute, Moscow, Russia$^{ e}$ \\
 $ ^{25}$ Max-Planck-Institut f\"ur Physik, M\"unchen, Germany \\
% ...
$ ^{29}$ Faculty of Science, University of Montenegro,
          Podgorica, Montenegro$^{ e}$ \\
% ...
$ ^{32}$ Dipartimento di Fisica Universit\`a di Roma Tre
          and INFN Roma~3, Roma, Italy \\
% ...
$ ^{49}$ Also at National Institute for Physics and Nuclear Engineering, Magurele, Bucharest, \linebreak Romania \\

\smallskip
 $ ^{\dagger}$ Deceased \\

\bigskip \noindent
% ...
$ ^c$ Supported by FNRS-FWO-Vlaanderen, IISN-IIKW and IWT
      and  by Interuniversity
Attraction Poles Programme,
      Belgian Science Policy \\
% ...
$ ^e$ Supported by the Deutsche Forschungsgemeinschaft \\
% ...
$ ^k$ Partially Supported by Russian Foundation
      for Basic Research,  grants  03-02-17291
      and  04-02-16445 \\
% ...
}

\newpage
\section{Introduction}
\label{intro}
\noindent

In the Standard Model (SM) the particle interactions conserve 
lepton flavour, although there is no underlying symmetry supporting this 
feature. However, experimental evidence for lepton flavour
violation (LFV) in 
solar and atmospheric neutrino oscillations has
been reported~\cite{SNO,SUPK}. The experimental upper
bounds on neutrino masses imply very small LFV effects in the charged
lepton sector. The observation of such effects would clearly 
indicate new phenomena beyond the SM.
\par
In $ep$ collisions at HERA, LFV processes $ep \rightarrow \mu X$ or $ep
\rightarrow \tau X$ lead to final states with a muon
or a tau and a hadronic system $X$. 
The LFV process can proceed via the exchange of a 
leptoquark (LQ), a boson with both lepton and baryon quantum number which 
appears naturally as a colour triplet scalar 
or vector boson in many extensions of the SM such as grand unified 
theories~\cite{GUT}, supersymmetry~\cite{SUSY}, 
compositeness~\cite{COMP} and technicolor~\cite{TECH}.
\par
In this paper a search for LFV phenomena is performed in $ep$ collision data
recorded during the years 1998-2000 by the H1 experiment, corresponding to an 
integrated luminosity of $66.5\,{\rm pb}^{-1}$ for $e^+p$ collisions and 
$13.7\,{\rm pb}^{-1}$ for $e^-p$ collisions at a centre-of-mass energy $\sqrt{s} = 319\,{\rm GeV}$.
The present results supercede those derived in previous searches at the H1
experiment using $e^+p$ collisions at $\sqrt{s} = 300\,{\rm GeV}$ \cite{H1OLDLFV}. 

\section{Experimental conditions}
\label{exp}

A detailed description of the H1 detector can be found in~\cite{detect}. 
In the following, only the detector components relevant for this analysis are 
briefly discussed. The origin of the H1 coordinate system is the nominal $ep$ 
interaction point, with the direction of the proton
beam defining the positive $z$-axis (forward direction). Transverse momenta
and azimuthal angles 
are measured in the $xy$ plane. The pseudorapidity is related to the polar 
angle $\theta$ by $\eta=-\ln\tan(\theta/2)$.
\par
A tracking system consisting of central and forward detectors is used to measure
charged particle trajectories and to determine the interaction vertex. 
The central tracker is composed of two concentric cylindrical drift 
chambers providing full acceptance for particles in the range 
\mbox{$22^\circ\!<\!\theta\!<\!160^\circ$}, complemented by a silicon vertex detector~\cite{Pitzl:2000wz} covering the range \mbox{$30^\circ\!<\!\theta\!<\!150^\circ$}. Transverse momenta ($P_T$) are
determined in the central region from the curvature of the particle 
trajectories in a magnetic field of $1.15\,{\rm T}$ with an 
effective resolution of \linebreak \mbox{$\sigma(P_T)/P_T\simeq0.01\cdot P_T({\rm
    GeV})$}. The tracking is complemented in the forward region
\mbox{$7^\circ\!<\!\theta\!<\!25^\circ$} by a system of drift chambers
perpendicular to the beam axis.
\par 
With a polar coverage of 
\mbox{$4^\circ\!<\!\theta\!<\!154^\circ$} and full azimuthal acceptance, the
liquid Argon (LAr) calorimeter encloses the tracking
chambers. It consists of an inner electromagnetic part with a
fine granularity and an outer hadronic part 
with a coarser granularity. 
The energy resolution of the LAr calorimeter for electrons 
and hadrons was determined in test beam measurements to 
be \mbox{$\sigma/E=12\%/\sqrt{E({\rm GeV})}\oplus1\%$} and 
\mbox{$\sigma/E=50\%/\sqrt{E({\rm GeV})}\oplus2\%$}, respectively \cite{TESTBEAM}.
In the backward region \mbox{$153^\circ\!<\!\theta\!<\!178^\circ$}, the LAr calorimeter is 
complemented by a lead-scintillating fibre spaghetti calorimeter.
\par 
The iron return yoke of the magnet is instrumented with streamer tubes to 
identify muon tracks. Further chambers of the central muon system are
positioned around the yoke to provide a precise muon track
measurement in the polar range \mbox{$5^\circ\!<\!\theta\!<\!175^\circ$}. 
Additional drift chambers positioned at either side of a toroidal magnet 
are employed to detect muons in the 
forward direction (\mbox{$3^\circ\!<\!\theta\!<\!17^\circ$}). 
\par
The luminosity is determined from the rate of the Bethe-Heitler process 
$ep\!\rightarrow\!ep\gamma$, measured using a photon detector located close 
to the beam pipe at \mbox{$z=-103\,\rm{m}$}. 
\par
Electrons are identified as compact and isolated electromagnetic clusters in
the calorimeter. 
Within the acceptance of the tracking detectors, an associated track is
required. A muon candidate is identified by associating an isolated 
track in the forward muon system or in the inner tracking system with a 
track segment or an energy deposit in the instrumented iron. The 
hadronic final state is reconstructed from the deposits in the LAr 
calorimeter in combination with tracking information. The hadrons are 
then combined into jets using the inclusive \linebreak $k_T$-algorithm \cite{ktjet} with a 
$P_T$-weighted recombination scheme where jets are treated as massless
and the separation parameter is set to one.

\section{LFV phenomenology and 
SM background processes}
\label{phen}

The LFV processes 
\mbox{$ep\rightarrow \mu X$} and \mbox{$ep\rightarrow \tau X$} 
can be attributed to LQs produced at HERA predominantly by electron-quark 
fusion.
\par
\begin{figure}
  \begin{center}
  \epsfig{file=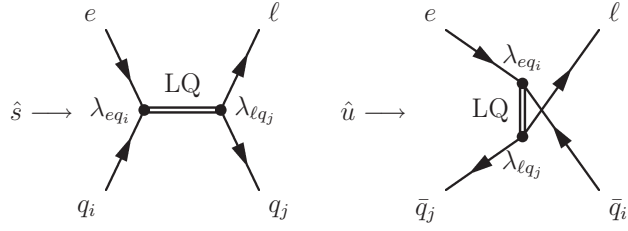,width=9cm}
    \caption[Feynman graphs of  $s$-channel resonant LQ production and 
    $u$-channel exchange of a LQ.]
    {
      Left: $s$-channel resonant LQ production and decay to a lepton-quark
      pair. Right: $u$-channel exchange of a LQ. The indices $i$ and $j$ 
      represent quark generation indices, such that $\lambda_{eq_i}$ denotes 
      the coupling of an electron to a quark of generation $i$, and 
      $\lambda_{\ell q_j}$ is the coupling of the outgoing lepton $\ell$ to a 
      quark of generation $j$. For $\ell=\mu,\tau$, the LQ introduces LFV.
    }
    \label{lqfeynman}
  \end{center}
\end{figure}

In the framework of the Buchm\"uller-R\"uckl-Wyler (BRW) 
effective model \cite{BRW}, LQs are classified into 14 types with 
respect to the quantum numbers spin, isospin and chirality. Leptoquarks 
carry both lepton ($L$) and baryon ($B$) quantum numbers. The 
fermion number \mbox{$F\!=\!L\!+\!3\,B$} is assumed to be 
conserved, taking values of $F\!=\!2$ for $e^-q$ processes 
and $F\!=\!0$ for $e^+q$ processes. Leptoquark processes proceed via $s$ channel
resonant LQ production or $u$ channel virtual LQ exchange, as shown in figure \ref{lqfeynman}. 
For LQ masses $m_{\rm LQ}$ well below the $e^\pm p$ centre-of-mass energy, the $s$ channel 
production of $F\!=\!2$ ($F\!=\!0$) LQs in $e^-p$ ($e^+p$) collisions 
dominates. For LQ masses 
above $319\,{\rm GeV}$ the
$s$ and $u$ channel processes become of equal importance and both $e^-p$ and $e^+p$ collisions
have similar sensitivity to virtual effects from $F\!=\!2$ LQs as well as from
$F\!=\!0$ LQs.
\par 
The BRW model assumes lepton flavour conservation (LFC) such that the LQs produced in
$ep$ collisions decay only to $eX$ or $\nu_eX$ final states. These LQs are 
referred to in the following as first generation LQs and have been studied in a 
recent H1 publication \cite{h1lq}. A general extension of the BRW model 
allows for the decay of LQs to final states containing a lepton of a different flavour, 
i.e. $\mu$ or $\tau$, and a jet, as illustrated in figure \ref{lqfeynman}. 
Non-zero couplings 
$\lambda_{eq_i}$ to an electron-quark pair and $\lambda_{\mu q_j}$ 
($\lambda_{\tau q_j}$) to a muon(tau)-quark pair are assumed. The 
indices\footnote{In the following the quark generation indices are attached 
only when it is relevant.} $i$ and $j$ represent quark generation indices, 
such that $\lambda_{eq_i}$ denotes the coupling of 
an electron to a quark of generation $i$, and $\lambda_{\ell q_j}$ is the coupling 
of the outgoing lepton $\ell$ to a quark of generation $j$.
\par
The double differential cross section for the $s$-channel tree level process 
is \cite{BRW}:
\begin{equation}
\frac{d^2\sigma_{\rm
    s}}{dxdy}=\underbrace{\frac{1}{32\pi\hat{s}}\vphantom{\frac{\lambda^2_{eq_i}}{(\hat{s}^2-m^2_{{\rm LQ}})^2}}}_
{\text{phase space}}\cdot
\underbrace{\frac{\lambda^2_{eq}\lambda^2_{\ell q}\hat{s}^2}{(\hat{s}^2-m^2_{{\rm LQ}})^2+m^2_{{\rm LQ}}\Gamma^2_{{\rm LQ}}}}_
{\text{Breit-Wigner {\rm LQ} propagator term}}\cdot 
\underbrace{q_i(x,\hat{s})\vphantom{\frac{\lambda^2_{eq_i}}{(\hat{s}^2-m^2_{{\rm LQ}})^2}}}_
{\text{parton density}}\times 
  \begin{cases}
    \frac{1}{2} & \text{scalar LQ}\\
    2(1-y)^2 & \text{vector LQ}~,
  \end{cases}
\label{eqExactSigmaschan}
\end{equation}
where $x$ is the Bj{\o}rken scaling variable, $y$ denotes the inelasticity of
the $ep$ scattering process, $\hat{s}=sx$ represents the square of the $eq$ centre-of-mass
energy and $\Gamma_{\rm LQ}$ is the total LQ width. A similar expression holds
for the $u$ channel exchange \cite{BRW}.
\par
An overview of the extended effective model
for the LQ coupling to $u$ and $d$ quarks is given in 
table~\ref{lqBRW}. For convenience only one LFV transition is considered:
either between the first and the second generations or between the first and
the third generations. The branching ratio $LQ\rightarrow\mu(\tau)q$ is given by
\begin{equation}
{\rm BR}=\beta_\ell\times \beta_{LFV}~~\text{with}
~~\beta_{LFV}=\frac{\Gamma_{\mu(\tau)q}}{\Gamma_{\mu(\tau)q}+\Gamma_{e}}~~\text{and}
~~\Gamma_{\ell q}=m_{{\rm LQ}}\lambda^2_{\ell q}\times 
  \begin{cases}
    \frac{1}{16\pi} & \text{scalar LQ}\\
    \frac{1}{24\pi} & \text{vector LQ}
  \end{cases}
\label{eqBranching}
\end{equation}
where \mbox{$\Gamma_{\ell q}$} denotes the partial LQ decay width to a lepton
\mbox{$\ell=e,\mu,\tau$} and a quark $q$ and where 
\mbox{$\beta_\ell\!=\!\Gamma_{\ell q}/(\Gamma_{\ell q}+\Gamma_{\nu_\ell q})$} is the fraction of 
decays into charged leptons. Some 
LQs, namely $S_0^L$, $S_{1}^L$, $V_0^L$ and $V_{1}^L$, can decay to a 
neutrino-quark pair resulting in $\beta_\ell=0.5$. Since neutrino flavours
cannot be distinguished with the H1
experiment, such final states are not covered in this search, but they 
are implicitly included in the search for first generation LQs \cite{h1lq}. 
\par
To determine the signal detection efficiencies, events with LQs are
generated using the LEGO \cite{lego} event generator with the CTEQ5L 
parametrisation of the parton distribution functions (PDF) of the proton \cite{CTEQ}. 
The LQ signal expectation is a function of the 
LQ type, mass, coupling constant and $\beta_{\rm LFV}$. The analysis usually 
requires a large number of simulated signal Monte Carlo (MC) samples. 
To overcome this technical difficulty, the LEGO program is used to produce a 
high statistics MC signal event sample generated according to a
double-differential cross section $d^2\sigma_{\rm generic}/(dx\,dQ^2)$ obtained
from \eqref{eqExactSigmaschan} by replacing the Breit-Wigner LQ propagator
term with a constant. This unique MC sample is used to calculate the
efficiency to select a LQ of a given type, mass $m_{\rm LQ}$, coupling $\lambda_{eq}$ and
$\beta_{\rm LFV}$ by attributing to each event a weight:
\begin{equation}\label{eqWeighting}
  w({\rm type},m_{\rm LQ},\lambda_{eq},\beta_{\rm LFV},x,Q^2)=
  \frac{
    \frac{d^2\sigma_{\rm exact}}{dxdQ^2}({\rm type},m_{\rm LQ},\lambda_{eq},\beta_{\rm LFV},x,Q^2)}
  {\frac{d^2\sigma_{\rm generic}}{dxdQ^2}(x,Q^2)}~,
\end{equation}
where $Q^2=sxy$ refers to the generated negative momentum
transfer squared and $x$ is the Bj{\o}rken scaling variable known at the generator
level. This procedure provides an exact prediction over 
the full range of LQ production parameters and avoids approaches like the
narrow width approximation or the high mass (contact interaction) approximation.
\par
The LQ kinematics are reconstructed using 
the double angle method~\cite{DAmeth}. The direction of the detected lepton
and jet are used to reconstruct the Bj{\o}rken scaling variable $x$ and therefore 
the LQ mass $m_{\rm LQ}^{\rm rec}=\sqrt{xs}$.
\par
The contributions from Standard Model (SM) background 
processes which may mimic the signal include neutral current (NC) and charged
current (CC) deep-inelastic scattering (DIS), photoproduction, lepton 
pair production and real W boson production. These processes are briefly 
described below:
\par
\begin{itemize}
\item {\bf NC DIS (\boldmath{$ep\rightarrow eX$})}\\
NC DIS processes contribute to the selected event sample if the scattered
electron is attributed to the tau electronic decay or if it is misidentified
as a narrow jet corresponding to a tau decay to hadrons. 
The NC DIS background is modelled using the event 
generator RAPGAP \cite{RAPGAP}. The proton PDFs are parametrised using 
CTEQ5L \cite{CTEQ} and hadronisation is performed using JETSET \cite{JETSET} 
parton showers and the Lund string fragmentation.
\item {\bf CC DIS (\boldmath{$ep\rightarrow \nu X$})}\\
Lepton flavour violating processes usually exhibit an imbalance in the measured calorimetric transverse
momentum due to either the presence of a minimally ionising muon in $\mu X$
final states or the escaping neutrino(s) from tau decays in $\tau X$ events. This imbalance is
exploited in the event selection. The CC DIS process leads to events with
genuine missing transverse momentum and therefore contributes to the selected
sample if hadrons or photons from the final state are misidentified as muons
or if tau decays are falsely reconstructed. The CC DIS contribution
is modelled using the DJANGO event generator \cite{DJANGO}. 
\item {\bf Photoproduction (\boldmath{$\gamma p\rightarrow X$})}\\
Events from photoproduction processes may 
contribute to the final selection if a hadron is wrongly
identified as a muon or if a narrow hadronic jet fakes the tau signature. 
This contribution is calculated using the event generator PYTHIA \cite{PYTHIA}. 
CTEQ5L \cite{CTEQ} serves as the 
proton PDF parametrisation and the photonic parton distribution parametrisation
GRV-LO \cite{GRVLO} is used. As PYTHIA only contains leading order
$2\!\rightarrow\!2$ processes, the multi-jet production cross section is 
underestimated \cite{H1JETSPHOTO}. Therefore, the prediction is scaled up by a
factor $1.2$ in this analysis, in agreement with previous analyses of jets in 
photoproduction \cite{H1JETSPHOTO}.
\item {\bf Lepton-pair production (\boldmath{$ep\rightarrow e\ell^+\ell^-X$})}\\
Lepton-pair production events contribute to the background because they may lead to high 
momentum leptons in the final state. In particular, inelastic di-muon events 
with one unidentified muon may fake the $\mu X$ LFV signature. The background 
samples include $ee$, $\mu\mu$ and $\tau\tau$ production generated with 
the event generator GRAPE \cite{GRAPE}.
\item $\mathbf{W}$ {\bf production (\boldmath{$ep\rightarrow eWX$})}\\
Real $W$ boson production leads to final states with isolated high $P_T$
leptons and missing transverse momentum. The simulated $W$ production samples are created with the
event generator EPVEC \cite{EPVEC} and include leptonic 
($e\bar{\nu}_e$, $\mu\bar{\nu}_\mu$, $\tau\bar{\nu}_\tau$) 
and hadronic $W$ decays.
\end{itemize}
\par
All signal and SM samples are passed through a detailed simulation
of the H1 detector \linebreak response based on the GEANT program \cite{GEANT} and the 
same reconstruction and analysis algorithms as used for the data.

\section{High \boldmath{$P_{T}$} Muon Signatures}
\label{muon}

Leptoquarks with couplings to the first and the second lepton generation can be
produced in $ep$ collisions and may decay to a muon and a quark. The
signature is an isolated high $P_T$ muon back-to-back to the hadronic system 
in the transverse plane. In general, a muon 
deposits a very small fraction of its energy in the LAr 
calorimeter. The signal is therefore expected to exhibit large $P_{T}^{\rm calo}$, 
which is the net transverse momentum reconstructed from all 
clusters recorded in the LAr calorimeter alone.
\par
The event preselection requires at least one muon with a transverse momentum above
$10\,{\rm GeV}$ in the polar angular range $10^{\circ}$ to
$140^{\circ}$ and at least one jet. The muon 
is required to be isolated. The
angular distance, $D=\sqrt{(\Delta\eta)^2+(\Delta\phi)^2}$, of the 
muon to the nearest track and to the nearest jet is required to be greater 
than $0.5$ and $1.0$, respectively. Only events with $P_{T}^{\rm calo}$
greater than $12\,{\rm GeV}$ are selected. In order to further exploit the event topology 
in the transverse plane, the cut $V_{ap}/V_p<0.3$ is employed, where
$V_{ap}/V_p$ is defined as the ratio of the anti-parallel to 
parallel projections of all energy deposits in the calorimeter 
with respect to the direction of $P_{T}^{\rm calo}$ \cite{H12001}.
\par
\begin{figure}
    \setlength{\unitlength}{1mm}
    \begin{picture}(138,185)
      \put(0,80){\includegraphics[width=0.49\textwidth]{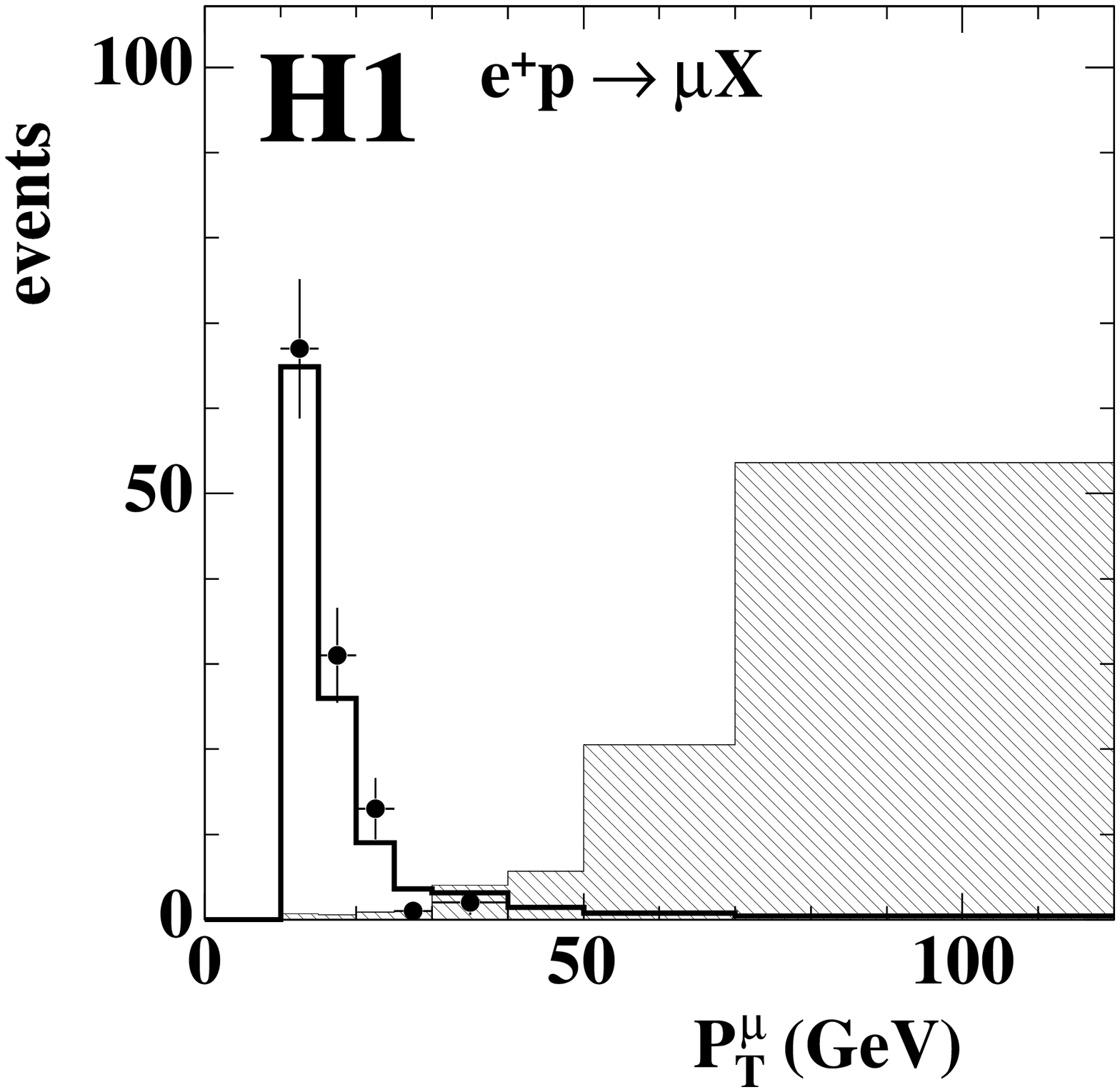}}
      \put(82,80){\includegraphics[width=0.49\textwidth]{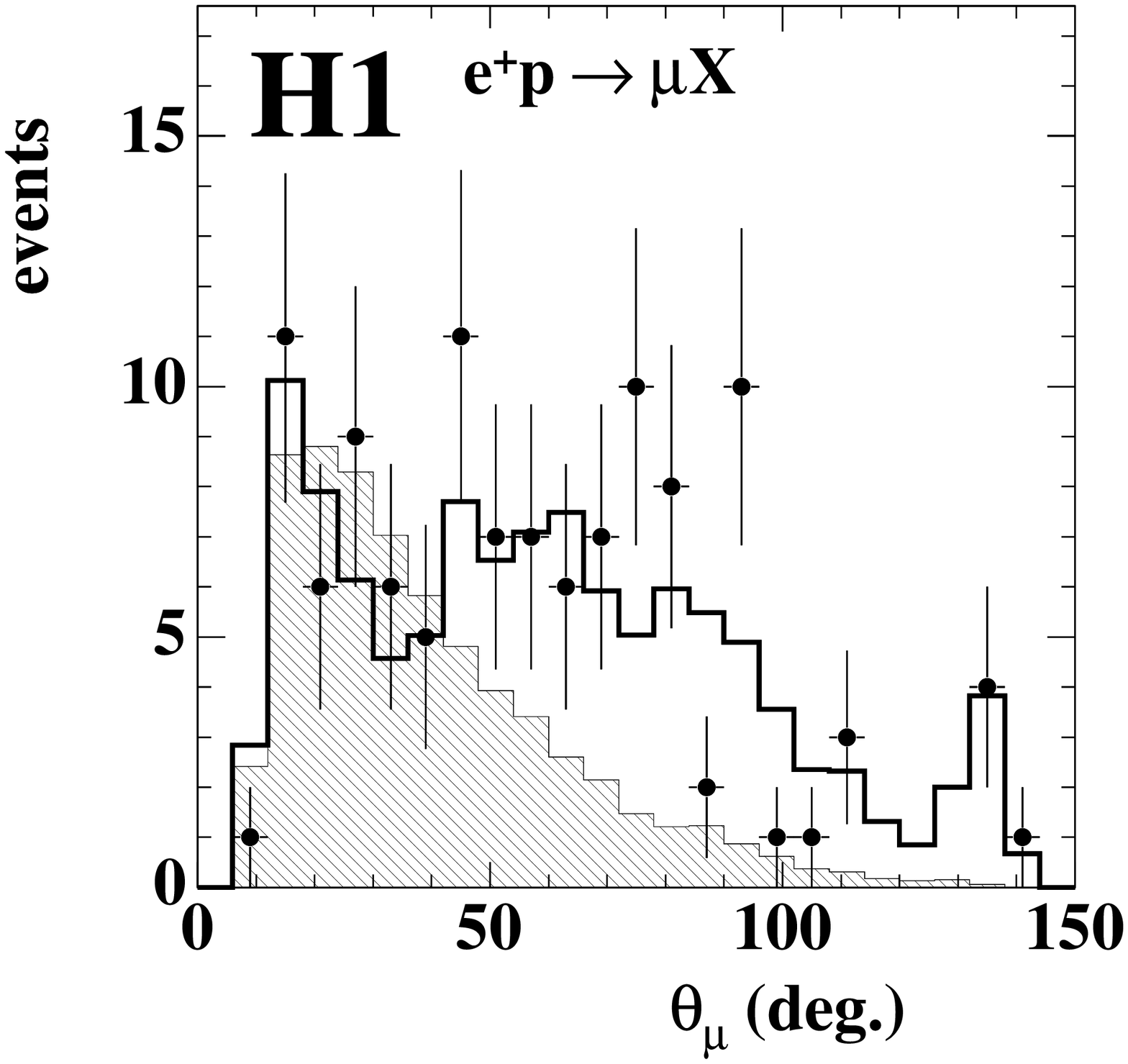}}
      \put(0,0){\includegraphics[width=0.49\textwidth]{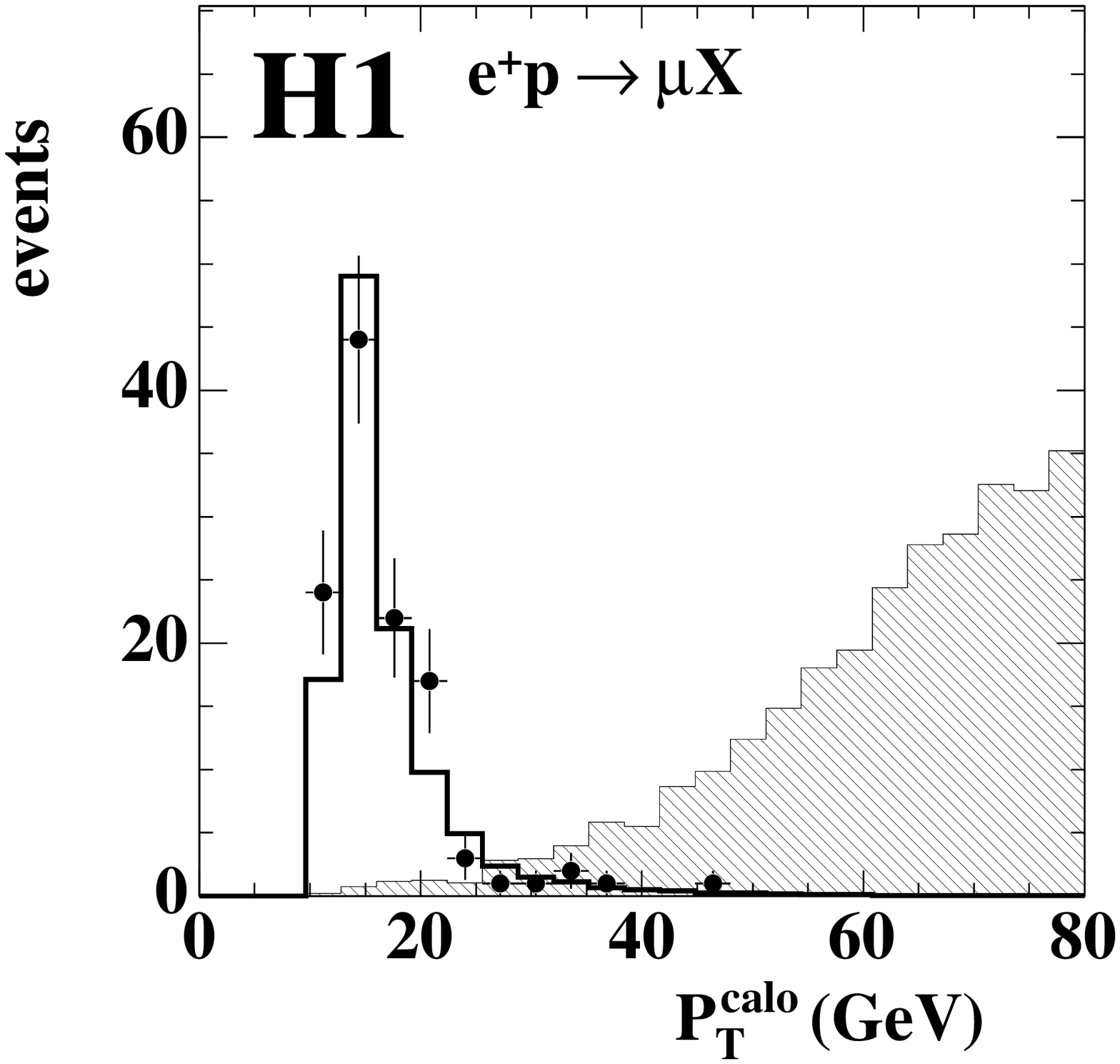}}
      \put(82,0){\includegraphics[width=0.49\textwidth]{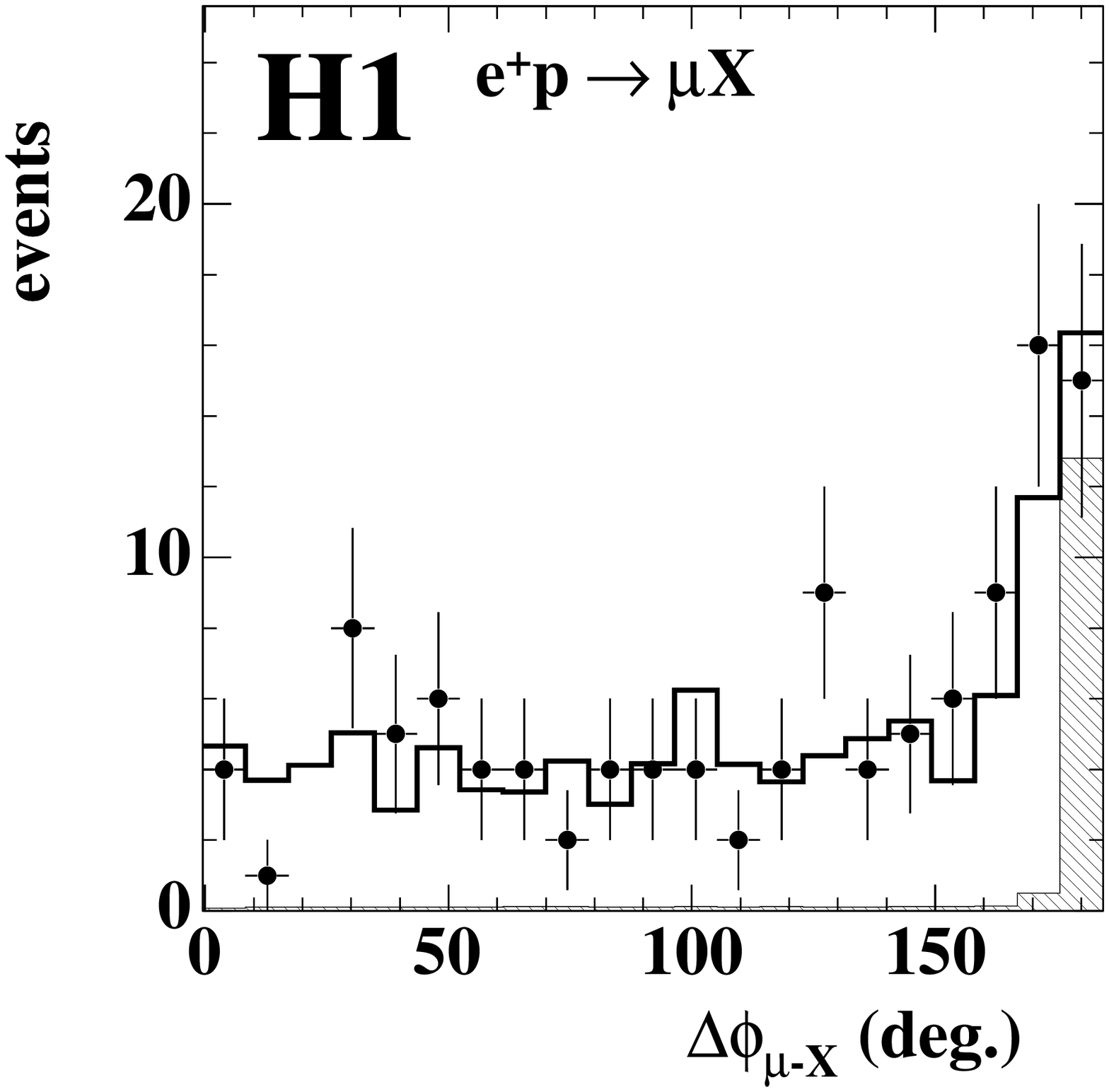}}
      \put(45,127){\epsfig{file=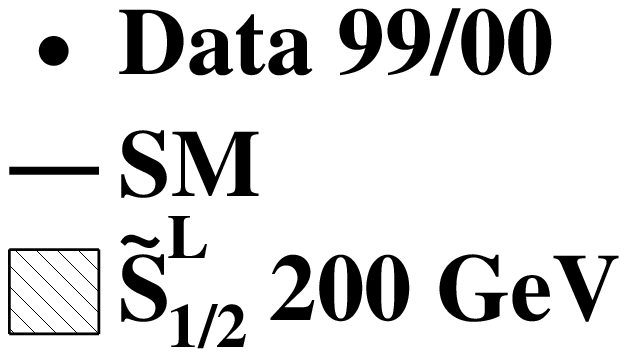,clip=,scale=0.40}}
      \put(46.90,137.7){\line(0,1){3.0}}
      \put(67,147){\small{\bf(a)}}
      \put(149,147){\small{\bf(b)}}
      \put(67,67){\small{\bf(c)}}
      \put(149,67){\small{\bf(d)}}
    \end{picture}
    \caption{
      Control distributions of the preselected $\mu X$ sample: 
      (a) muon transverse momentum, (b) muon polar angle, (c) transverse
      momentum as measured from the calorimeter deposits and (d) acoplanarity 
      between the muon and the hadronic final state $X$. Data (points) from
      $e^+p$ collisions are compared to the SM expectation (histogram). The LFV signal MC
      sample of a leptoquark $\tilde{S}^L_{1/2}$ with $m_{\rm LQ}=200\,{\rm GeV}$ and
      $\lambda_{eq}\!=\!\lambda_{\mu q}\!=\!0.3$ is shown hatched with arbitrary normalisation 
      in each plot.}
    \label{muonpreselplots}
\end{figure}

Figure \ref{muonpreselplots} displays the distributions of the transverse
momentum of the muon, its polar angle $\theta_\mu$, $P_{T}^{\rm calo}$ and the acoplanarity 
$\Delta\phi_{\mu-X}$ between the muon and the hadronic final state $X$ 
after the muon preselection. The data passing the preselection are well 
described by the SM prediction. The signal corresponding to a scalar LQ with 
\mbox{$m_{\rm LQ}=200\,{\rm GeV}$} is also shown. It displays muons with large $P_{T}^{\mu}$ produced
predominantly in the forward direction (low $\theta_\mu$) in events with
significant $P_{T}^{\rm calo}$ and back-to-back topology 
\mbox{$\Delta\phi_{\mu-X}\!\simeq\!180^\circ$}.
\par
In the final LFV selection step, the NC DIS background is further
suppressed by rejecting events with identified electrons, and by accepting only
events with an imbalance of the calorimeter deposits, 
\mbox{$P_{T}^{\rm calo}>25\,{\rm GeV}$}, and with a back-to-back topology, 
\mbox{$\Delta\phi_{\mu-X}\!>\!170^\circ$}. The latter selection criterion is
only applied for events for which the hadronic final state is well contained
in the detector, with the reconstructed polar angle \mbox{$7^\circ<\theta_X<140^\circ$}.
\par
The selection efficiency ranges from 40\% to 60\% depending on
the LQ mass and type (see table \ref{summarytable}). 

\section{High \boldmath{$P_{T}$} Tau Signatures}
\label{tau}

Leptoquarks with couplings to the first and the third lepton generation can be
produced in $ep$ collisions and may decay to a tau and a quark. Tau leptons
are identified using the electronic, muonic and hadronic decays of the tau. 

\subsection{Electronic tau decays}
\label{etau}

The final state resulting from the electronic tau decay, 
$\tau\rightarrow e\nu_e\nu_\tau$, leads to an event topology that is very
similar to that of high $Q^2$ NC DIS events. The preselection follows that presented in \cite{H1NCDIS}. 
A reconstructed jet with a minimal
transverse momentum of $P_T^j>25\,{\rm GeV}$ back-to-back in the transverse 
plane to an electron with $P_T^e>10\,{\rm GeV}$ is required. The kinematic 
domain is restricted to $Q^2>1000\,{\rm GeV^2}$ and $y>0.1$. Figure 
\ref{taupreselplots}(a) shows the distribution of $P_{T}^{\rm miss}$ after
this preselection, where $P_{T}^{\rm miss}$ is defined as the total missing
transverse momentum reconstructed from all observed particles.
\par
In the final selection a large missing transverse momentum $P_{T}^{\rm miss}>20\,{\rm GeV}$
is required in order to account for the expected missing momentum carried by
the neutrinos produced in the tau decay. These neutrinos are boosted 
along the electron direction, implying an imbalance between the transverse momenta 
of the electron $P_T^e$ and the hadronic final state $P_T^X$. Hence, 
the restriction $P_T^e/P_T^X<0.8$ further reduces NC DIS background. In addition, the 
azimuthal distance between the missing transverse momentum and the electron 
must not exceed $20^\circ$. The remaining NC DIS background, due to mismeasured electron 
energies leading to missing energy near the electron, is reduced by the requirement 
$P_T^{\rm e-clu}/P_T^{\rm e-trk}>0.7$, where $P_T^{\rm e-clu}$ is
measured from the electromagnetic cluster and $P_T^{\rm e-trk}$ from tracking 
information. Any events with additional isolated muons are excluded from the 
electronic tau decay channel. The final selection in the electronic tau decay channel yields an 
efficiency normalised to all tau decays of $3\%$ to $10\%$, 
which is limited by the branching fraction 
$BR(\tau\rightarrow e\nu_e\nu_\tau)=17.8\%$\,\cite{PDG} and 
dependent on the assumed LQ mass and type (see table \ref{summarytable}). 

\subsection{Muonic tau decays}
\label{mutau}

Muonic tau decays $\tau\rightarrow \mu\nu_\mu\nu_\tau$ result in 
similar final states as the high $P_T$ muon signatures described 
in section \ref{muon}. The same selection cuts described therein 
are applied here. To account for possible effects due to 
different muon kinematics resulting from a tau decay, 
the selection efficiency was studied in detail
with a LFV MC signal sample with a $\tau X$ final state and a subsequent muonic
tau decay. 
The selection efficiency ranges between 4\% and 8\%, which is
dependent on the LQ mass and type, normalised to all tau decays and limited by 
\mbox{$BR(\tau\rightarrow \mu\nu_\mu\nu_\tau)=17.4\%$}\,\cite{PDG} 
(see table \ref{summarytable}). 

\subsection{Hadronic tau decays}
\label{hadtau}

The hadronic decays of the high $P_T$ tau lead to a typical signature 
of a high $P_T$ ``pencil-like'' jet. The signal topology is a di-jet event 
with no leptons. The tau-jet is characterised by a narrow energy deposit 
in the calorimeter and low track multiplicity with predominantly one or three 
tracks in the identification cone of the jet. The neutrino from 
the tau decay are boosted along the direction of the hadrons. 
The missing transverse momentum in the event is aligned with the tau-jet. 
\par
Tau-jet candidates are defined as jets with exactly one or three tracks in the 
tau-jet cone with an opening angle that varies between $5^\circ$ and
$30^\circ$ with decreasing jet momentum. The tracks are required 
not to be associated with identified electrons or muons and the scalar sum of 
their transverse momenta is required to be larger than 2 GeV. The fine 
granularity of the LAr calorimeter is used to match extrapolated tracks with 
energy deposits in the calorimeter and to separate additional neutral particles
associated to the tau candidate from unmatched energy deposits in the tau-jet cone. The sum of
the four-vectors of the tracks and of the neutral particles defines the
tau-jet candidate four-vector. 
\par
\begin{figure}
    \setlength{\unitlength}{1mm}
    \begin{picture}(138,185)
      \put(0,80){\includegraphics[width=0.49\textwidth]{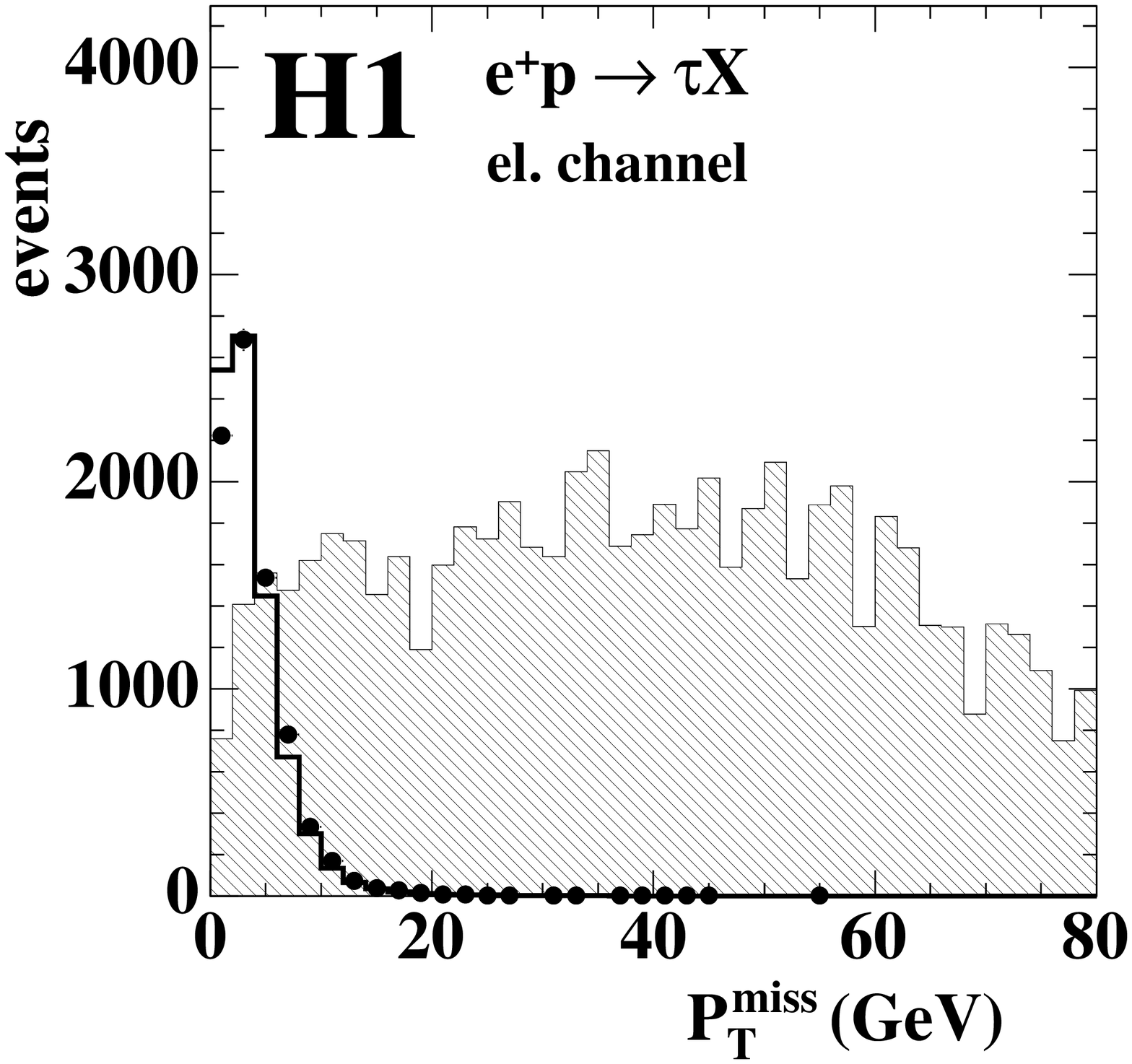}}
      \put(82,80){\includegraphics[width=0.49\textwidth]{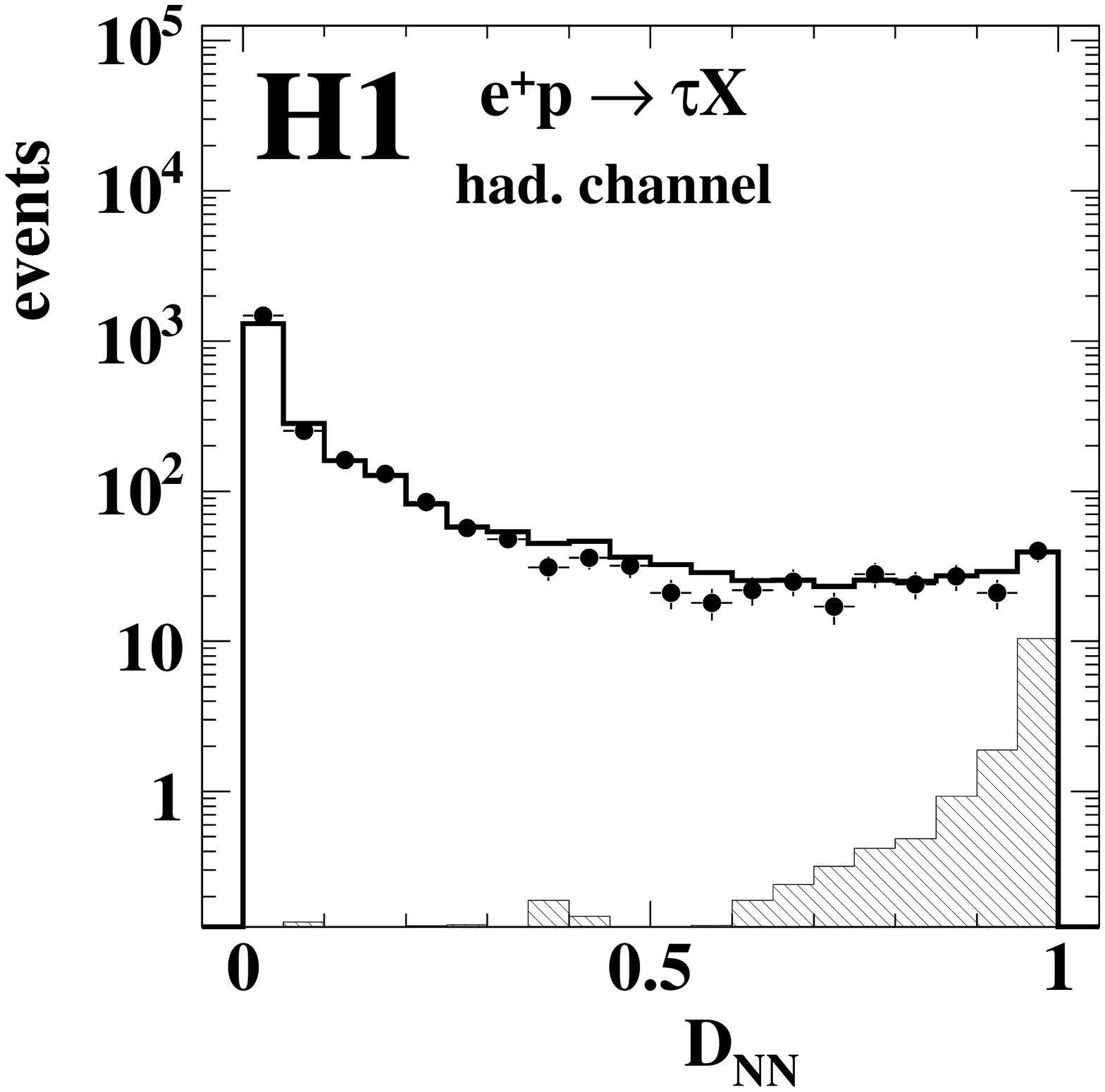}}
      \put(0,0){\includegraphics[width=0.49\textwidth]{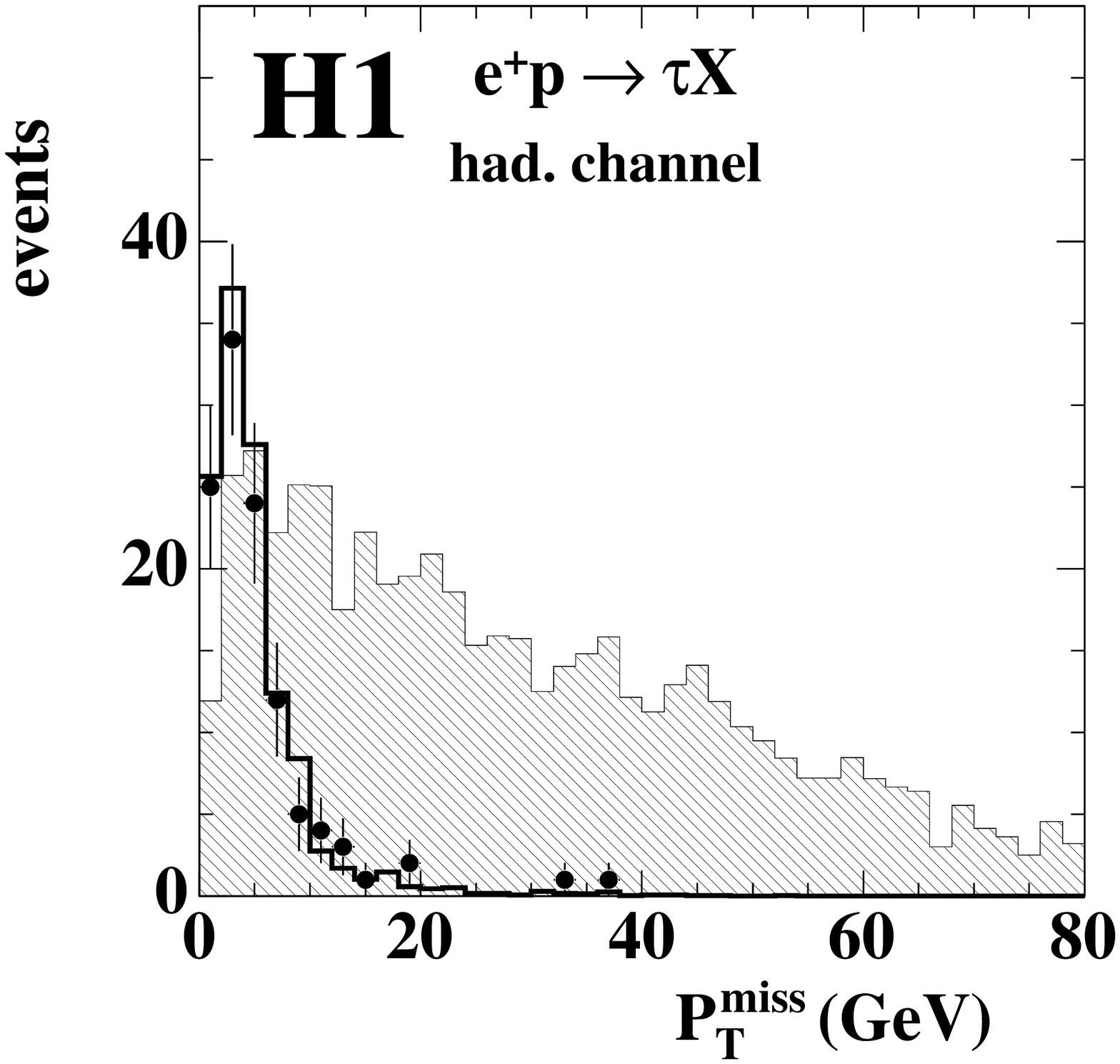}}
      \put(82,0){\includegraphics[width=0.49\textwidth]{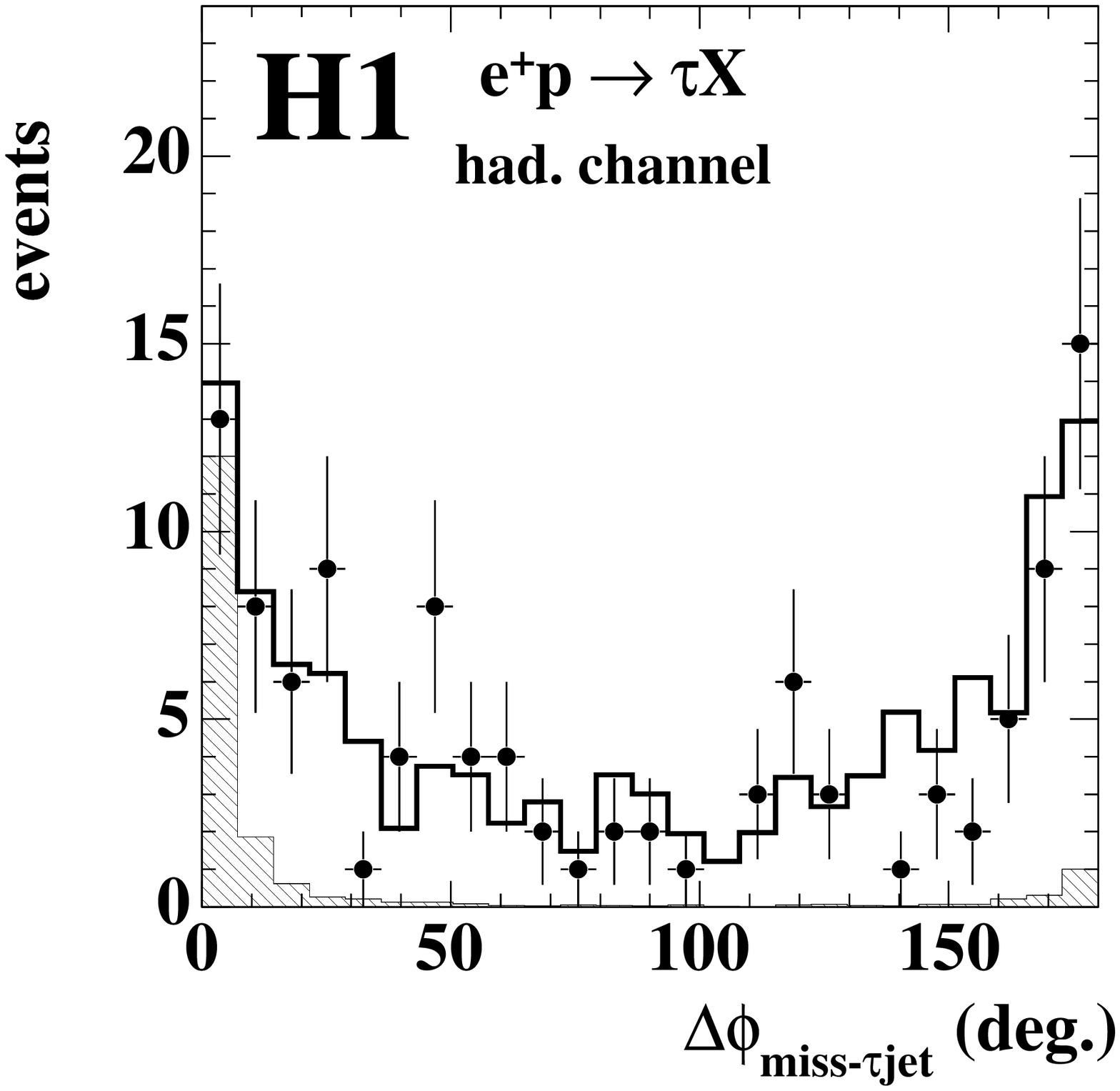}}
      \put(45,124){\epsfig{file=legend_small.eps,clip=,scale=0.40}}
      \put(46.9,134.7){\line(0,1){3.0}}
      \put(67,147){\small{\bf(a)}}
      \put(149,147){\small{\bf(b)}}
      \put(67,67){\small{\bf(c)}}
      \put(149,67){\small{\bf(d)}}
      \put(37,55){\small{${\cal D}_{\rm NN}>0.8$}}
      \put(119,55){\small{${\cal D}_{\rm NN}>0.8$}}
    \end{picture}
    \caption{
      Distributions of the preselected $\tau X$ sample: 
      (a) missing transverse momentum in the 
      electronic tau decay channel and (b) neural net tau-jet discriminant 
      after the preselection in the hadronic tau decay channel. The restricted
      sample obtained after the additional cut ${\cal D}_{\rm NN}\!>\!0.8$ in
      the hadronic channel: (c) missing 
      transverse momentum and (d) acoplanarity between the tau-jet and 
      the missing transverse momentum. The LFV signal MC sample of a 
      leptoquark $\tilde{S}^L_{1/2}$ with $m_{\rm LQ}=200\,{\rm GeV}$ and 
      $\lambda_{eq}\!=\!\lambda_{\tau q}\!=\!0.3$ is shown hatched with arbitrary normalisation in 
      each plot.}
    \label{taupreselplots}
\end{figure}

In the preselection step at least two jets with 
a transverse momentum \mbox{$P_T^{\rm jet1}>20\,{\rm GeV}$} 
and \mbox{$P_T^{\rm jet2}>15\,{\rm GeV}$} reconstructed in the polar angle 
range \mbox{$7^\circ<\theta_{\rm jets}<145^\circ $} are required. One jet must 
fulfil the criteria of a tau-jet candidate with $\theta_{\tau{\rm
    jet}}>20^\circ$. In addition, the calorimetric shower
shape and tracking signature are exploited to validate the tau-jet candidates. The
following estimators are used to separate a tau-jet from quark or
gluon induced jets: the number of all tracks associated to the tau-jet candidate, 
the distance in $\eta-\phi$ between tracks 
and calorimetric clusters, the number of calorimeter cells of the tau-jet
$n_{\rm cells}$, the radial extension of the calorimetric deposits 
$\langle r \rangle=\sum_{i=1}^{n_{\rm cells}}E_ir_i/\sum_i E_i$, the standard
deviation $\sigma(r)=\sqrt{\langle r^2 \rangle - \langle r \rangle ^2}$ and
the invariant tau-jet mass reconstructed from calorimeter cells. A 
neural net algorithm is employed and trained using the six estimator
variables, as explained in \cite{H1TAULEPTONS}. The neural net yields
a discriminator variable ${\cal D}_{NN}$ in the range $0\le{\cal D}_{\rm
  NN}\le1$ with values close to $0$ for quark or gluon induced jets and 
close to $1$ for hadronic tau decays. The distribution of the discriminant ${\cal
  D}_{NN}$ after the preselection is depicted in figure
\ref{taupreselplots}(b). 
The distributions of 
$P_{T}^{\rm miss}$ and \mbox{$\Delta\phi_{\rm miss-\tau jet}$} after 
requiring ${\cal D}_{NN}>0.8$ are shown in figures 
\ref{taupreselplots}(c),(d). This requirement yields a signal efficiency of 
$80\%$ and a quark or gluon induced jet rejection of $95\%$. After all preselection criteria $16$ ($112$) events are selected in $e^-p$ ($e^+p$) 
data sample for $22.0\pm1.0(stat.)$ ($121.1\pm5.3(stat.)$)  expected from
the SM. 
\par
The final selection step in the hadronic tau decay channel makes use of the
characteristic large missing transverse momentum carried by the tau neutrino 
which is expected to be in the direction of the tau-jet. The
difference in $\phi$ between the missing transverse momentum vector and the
tau-jet, \mbox{$\Delta\phi_{\rm miss-\tau jet}$}, is required to be below
$20^\circ$.  A minimal value of 
\mbox{$P_{T}^{\rm miss}> 12\,{\rm GeV}$} is chosen for an 
accurate determination of the direction. In addition 
\mbox{$P_{T}^{\rm calo}> 12\,{\rm GeV}$} is required. 
The final signal selection efficiency in the hadronic tau decay channel 
varies between 3\% and 13\%, normalised to all tau decays and limited by the branching
  fraction $BR(\tau^-\rightarrow \nu_\tau+{\rm hadrons})=64.8\%$\,\cite{PDG} 
(see table \ref{summarytable}). 


\section{Systematic Uncertainties}
\label{sys}

The following experimental systematic uncertainties are considered:
\begin{itemize}
\item
  The energy of electrons is measured with a systematic uncertainty in the
  range from $0.7\%$ to $3\%$ depending on the polar angle. The uncertainty of
  the electron direction is estimated to be less than $3\,{\rm mrad}$ in
  $\theta$ and $1\,{\rm mrad}$ in $\phi$.
\item
  The scale uncertainty on the transverse momentum of high $P_T$ muons amounts 
  to $5\%$. The uncertainty on the reconstruction of the muon direction is 
  $3\,{\rm mrad}$ in $\theta$ and $1\,{\rm mrad}$ in $\phi$.
\item
  For the hadronic final state, an energy scale uncertainty of $2\%$ and
  a direction uncertainty of $20\,{\rm mrad}$ are assumed. 
\item
  The luminosity of the analysed datasets is known to $1.5\%$.
\end{itemize}

The effects of these systematic uncertainties on the signal and the expected SM background 
are evaluated by shifting the relevant quantities in the MC simulation by their uncertainty 
and adding all resulting variations in quadrature.
\par
Systematic errors accounting for normalisation uncertainties on the expected
background determined from the individual MC event generators are estimated to be 10\% 
for NC DIS and Lepton-pair production, 15\% for W production and 30\% for 
photoproduction and CC DIS. The relatively large error of 30\% on
photoproduction and CC DIS is due to uncertainties on higher-order
corrections. The errors associated to the background normalisation are added
in quadrature to the experimental error to calculate the total error of the SM 
prediction.
\par
The main theoretical uncertainty on the signal cross section originates from 
the parton densities. This uncertainty is estimated as described in 
\cite{h1lq}. It is found to be $5\%$ for LQs coupling to up-type quarks
and varies between $7\%$ at low masses and $30\%$ at masses around 
$290\,{\rm GeV}$ for LQs coupling to down-type quarks. The correlation
between different channels is taken 
into account for the statistical interpretation and limit calculation \cite{BOTJE}. A 
detailed description of the analysis can be found in \cite{LINUSPHD}.

\section{Results}
\label{res}

No candidate is found in the final data sample of the muon channel. 
The expected number of SM background events is $1.03\pm0.32$ in 
the $e^+p$ set and \mbox{$0.18\pm0.06$} in the $e^-p$ sample.
The largest contribution to this background comes from muon-pair 
production and the muonic decays of $W$ bosons. These results apply equally 
to the muonic tau decay channel.
\par
In the electronic tau decay channel no data event is found compared to a SM expectation
of $0.28\pm0.19$ events in the $e^-p$ sample and $1.24\pm0.55$ events in the $e^+p$ data. NC DIS events with a mismeasured electron energy are the largest background contribution.
\par
No $e^-p$ data event passes the final selection criteria in the hadronic tau decay
channel. The expected SM background amounts to $0.29\pm0.06$. 
One event is selected in the $e^+p$ data for an expected SM prediction of
$2.63\pm0.57$, dominated by NC DIS and photoproduction processes.
\par
The results of the final selection in all channels are summarised in 
table \ref{summarytable}. Typical signal selection efficiencies for some LQ 
types with a mass of $150\,{\rm GeV}$ and $500\,{\rm GeV}$ 
are also given. The observation is in agreement with the SM 
prediction and no evidence for LFV is found by the present analysis. Limits on 
the model parameters presented in section \ref{phen} are calculated as described
in the following section.

\section{Limits}
\label{limits}

The results of the search are interpreted in terms of exclusion limits on 
the mass and the coupling of LQs that may mediate LFV. 
The LQ production mechanism at HERA involves non-zero coupling to the first generation fermions
\mbox{$\lambda_{eq} > 0$}. The LFC decays \mbox{${\rm LQ}\!\rightarrow\!
eq$} or \mbox{${\rm LQ}\!\rightarrow\!\nu_eq$} are therefore
possible. In order to cover the full LQ decay width and 
to generalise the results of LFV searches in $ep$ collisions to an arbitrary
weight between the LFC and LFV decay channels, the searches for LFC decays
presented in \cite{h1lq} are combined with each of the 
LFV search channels $\mu X$ or $\tau X$ of the present analysis. It is assumed that only 
one of the couplings $\lambda_{\mu q}$ and $\lambda_{\tau q}$ is non--zero and therefore 
the $\mu X$ and $\tau X$  channels enter the limits calculation separately. 
A modified frequentist method with a likelihood ratio as the test statistic is used to combine 
the individual data sets and search channels \cite{HIGGSLEP}. 
\par
In first generation LQ signals are searched for in about 400 bins 
in the $m_{\rm LQ}\!-\!y$ plane and the observed data is in agreement 
with the irreducible SM NC and CC background \cite{h1lq}.
For the LFV channel $\mu q$ ($\tau q$), the couplings $\lambda_{e q}$ and
$\lambda_{\mu q}$ ($\lambda_{\tau q}$) and the LQ mass determine the total 
production cross section, which is compared to the selected data from the LFV search
channel and the first generation results. A combined test
statistic is built and used to set limits as a function of $\lambda_{e q}$,
$\lambda_{\mu q}(\lambda_{\tau q})$ and $m_{\rm LQ}$. This procedure
implicitly includes in the analysis the decays to a neutrino of any flavour
and a quark.
\par
\begin{figure}
  \setlength{\unitlength}{1mm}
  \begin{picture}(138,155)
    \put(0,80){\includegraphics[width=0.49\textwidth]{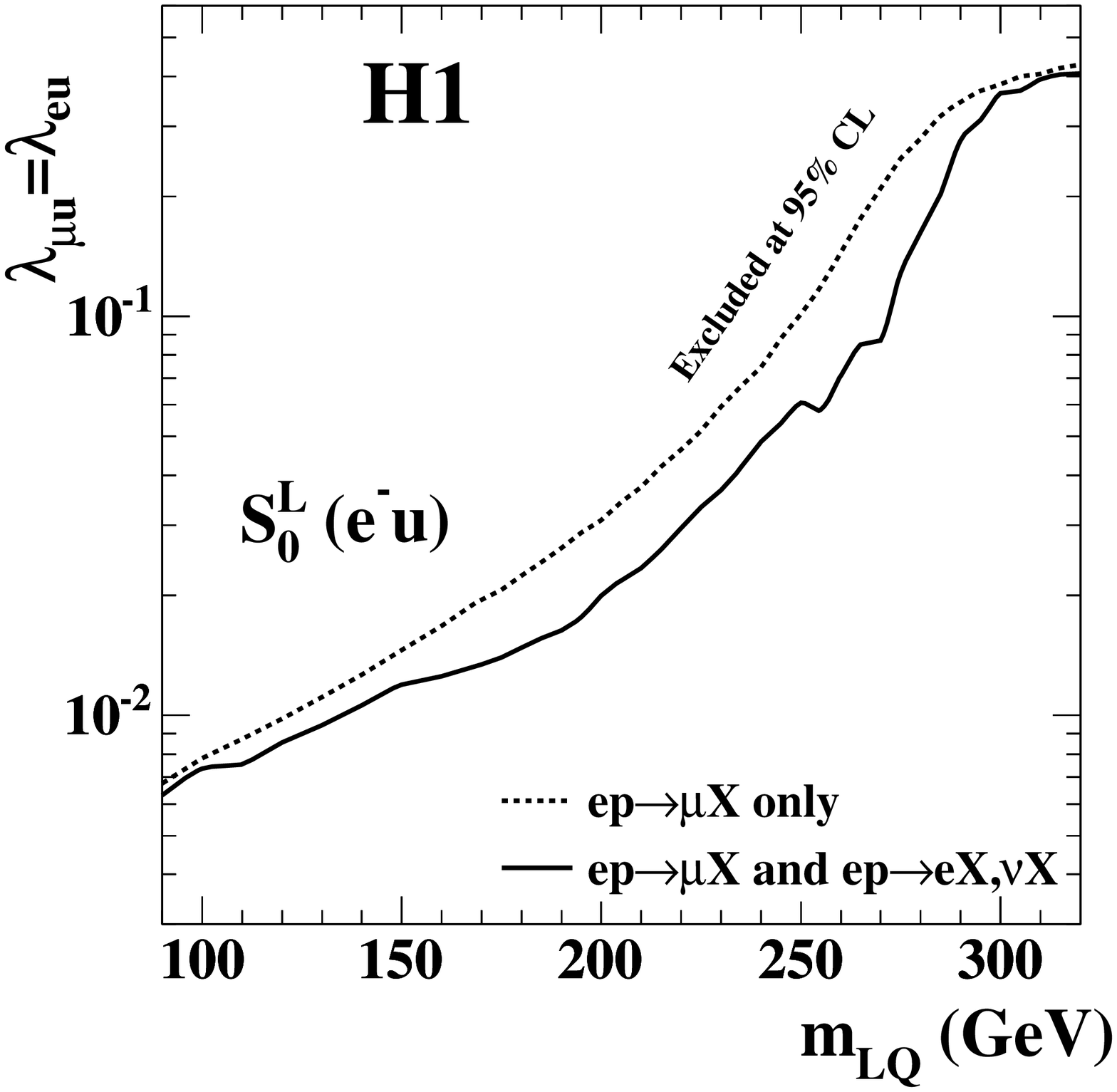}}
    \put(82,80){\includegraphics[width=0.49\textwidth]{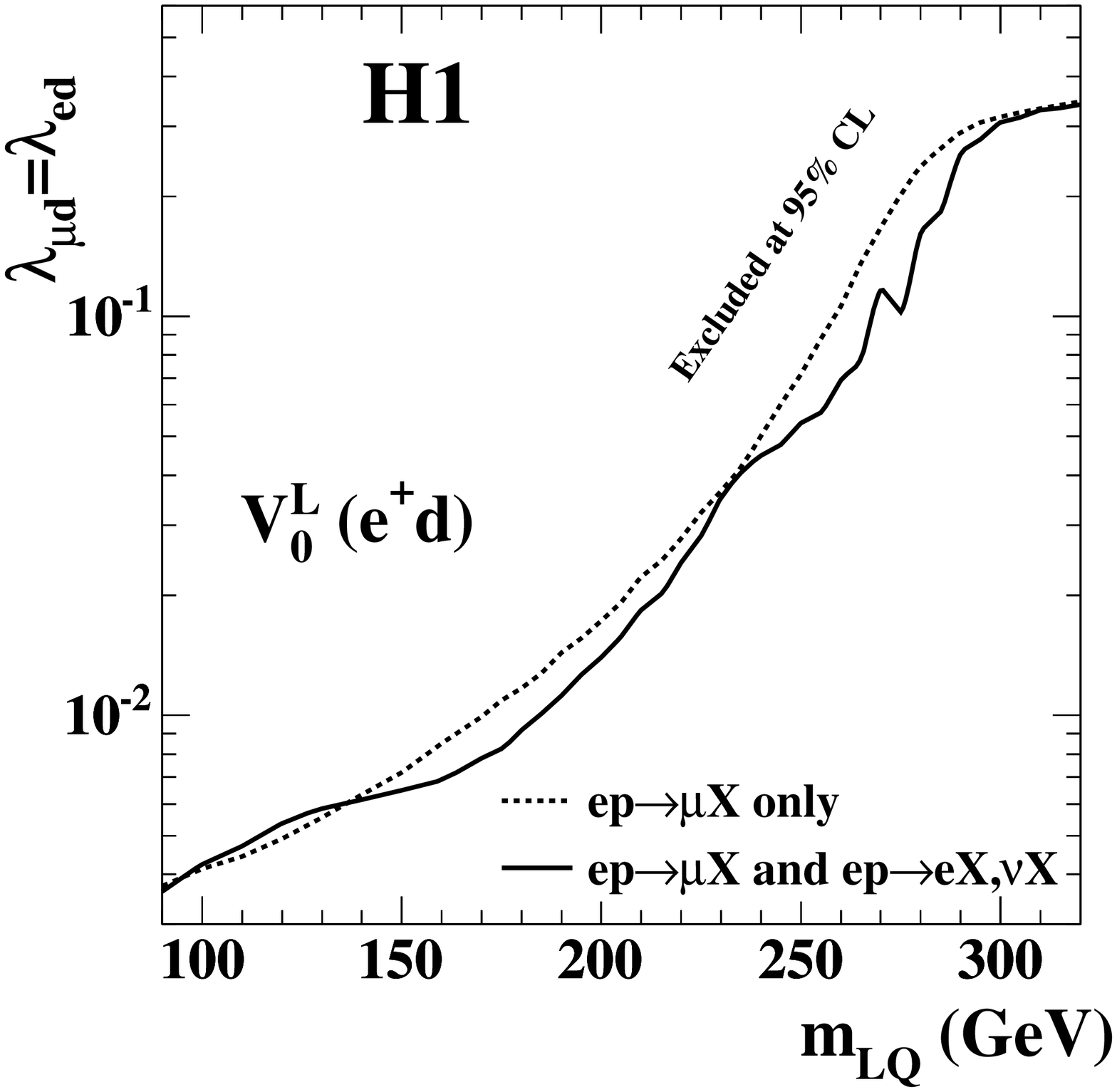}}
    \put(0,0){\includegraphics[width=0.49\textwidth]{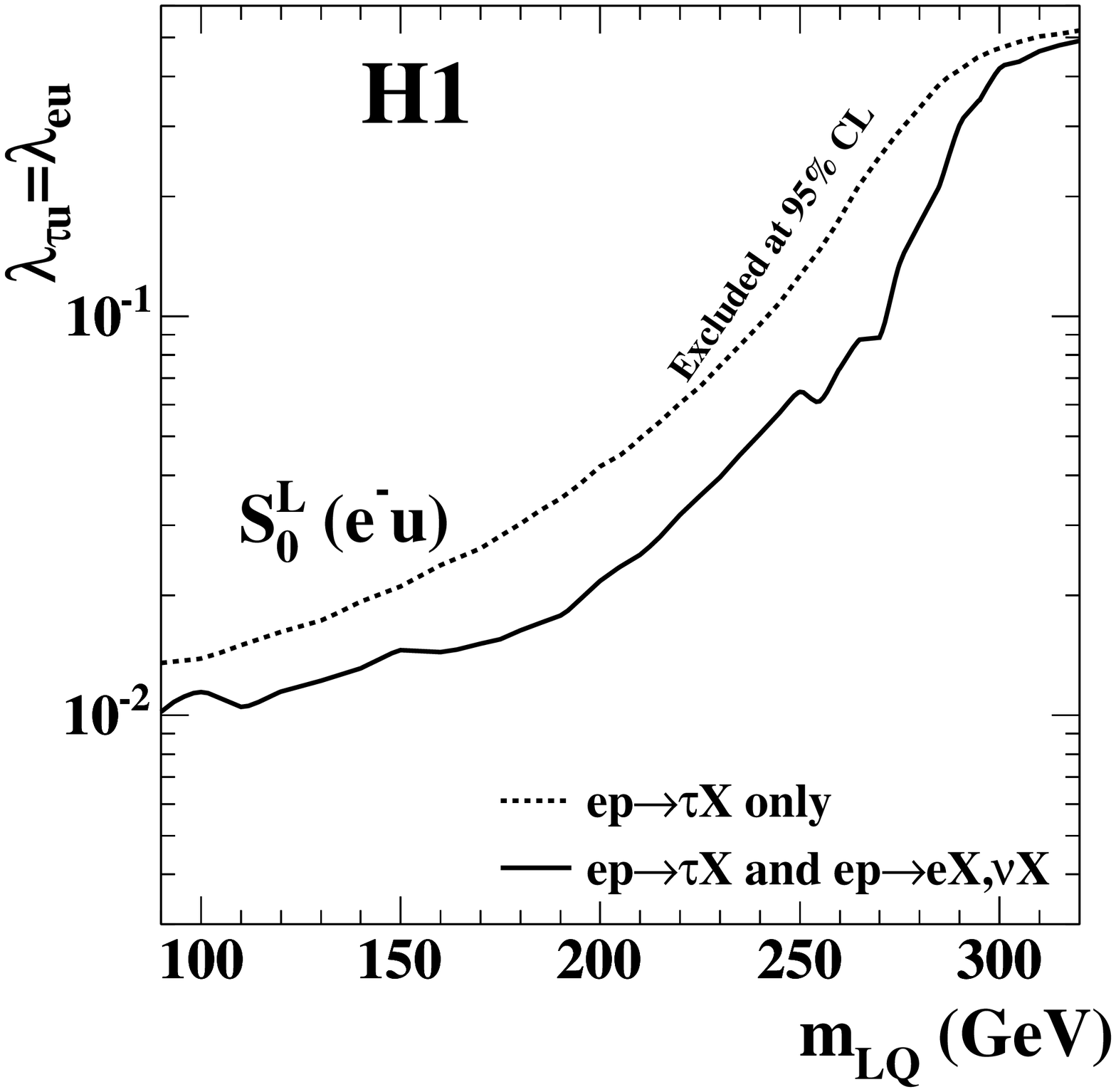}}
    \put(82,0){\includegraphics[width=0.49\textwidth]{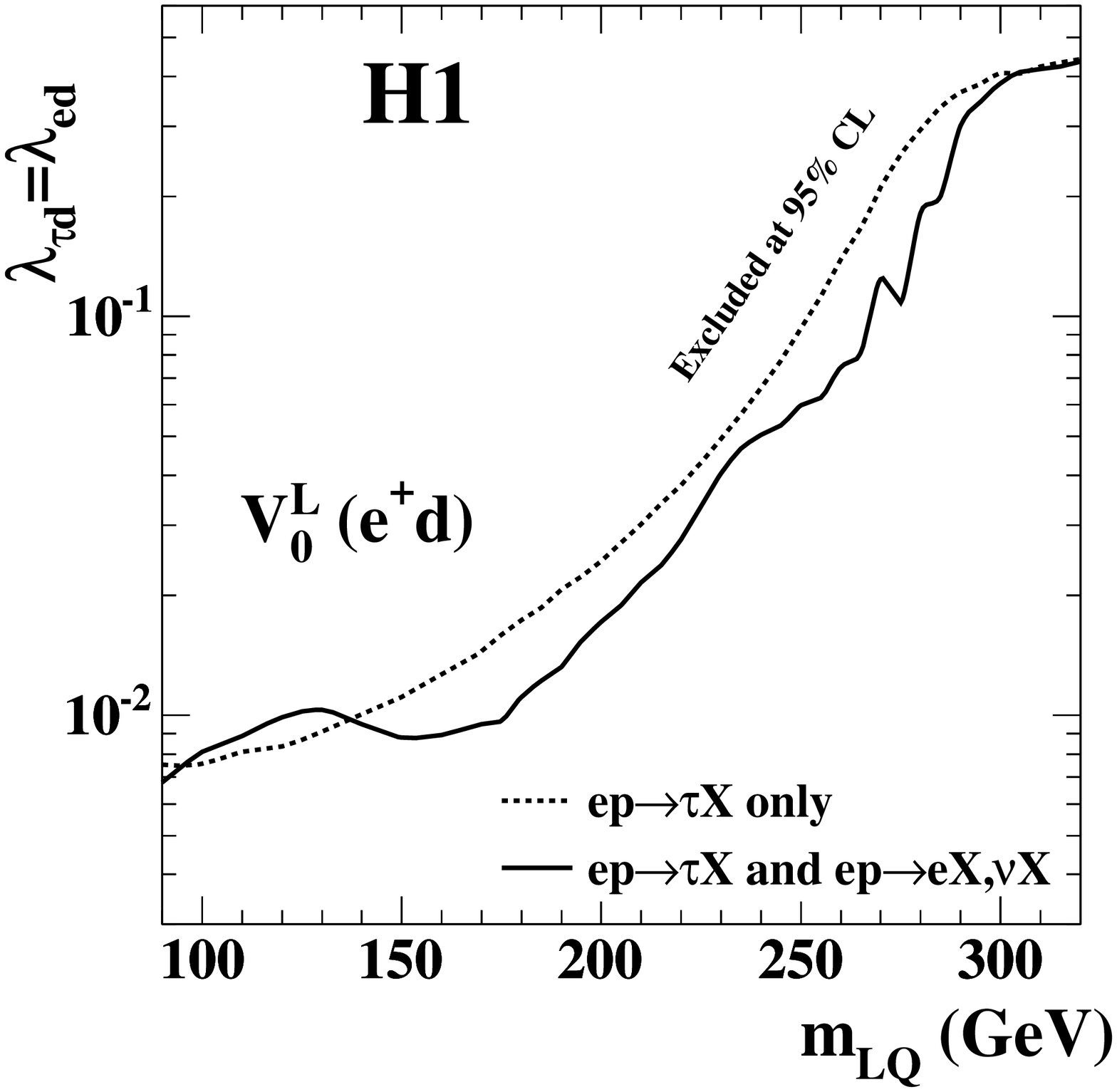}}
    \put(15,148){\small{\bf(a)}}
    \put(97,148){\small{\bf(b)}}
    \put(15,68){\small{\bf(c)}}
    \put(97,68){\small{\bf(d)}}
    \end{picture}
    \caption{
    Comparison of limits at 95\% CL on the coupling constants 
    $\lambda_{\ell q}$ under the assumption $\lambda_{\ell q}\!=\!\lambda_{eq}$
    as a function of the leptoquark mass $m_{\rm LQ}$ for: 
    (a) $S_0^L$ on $\lambda_{\mu u}\!=\!\lambda_{eu}$,
    (b) $V_0^L$ on $\lambda_{\mu d}\!=\!\lambda_{ed}$,
    (c) $S_0^L$ on $\lambda_{\tau u}\!=\!\lambda_{eu}$, and
    (d) $V_0^L$ on $\lambda_{\tau d}\!=\!\lambda_{ed}$.
    The areas above the dashed
    lines represent the exclusion regions using only the lepton flavour 
    violating leptoquark decay 
    channels $ep\!\rightarrow\!\mu X$ and $ep\!\rightarrow\!\tau X$, respectively.
    The limits after combination with the results of the search for first
    generation leptoquarks are shown as solid lines. 
    }
    \label{figlimitscombined}
\end{figure}

Figure \ref{figlimitscombined} shows limits before and after combination with
the search for first generation LQs for the LQ types $S_0^L$ and
$V_0^L$ up to LQ masses of $320\,{\rm GeV}$ assuming 
\mbox{$\lambda_{eq}\!=\!\lambda_{\mu q}$} and 
\mbox{$\lambda_{eq}\!=\!\lambda_{\tau q}$}, i.e. \mbox{$\beta_{\rm  LFV}\!=\!0.5$}, in
the resonance production region. 
The comparison for these types exemplifies that the limits on those LQs which can 
decay to a neutrino-quark pair, namely $S_0^L$, $S_{1}^L$, $V_0^L$ and
$V_1^L$, benefit most from the combination with the search for first 
generation LQs which covers decays to a neutrino-quark pair. 
In the high mass regime \mbox{$m_{\rm LQ}\!\gg\!\sqrt{s}$} (contact
interaction region) 
the obtained limits are similar to those deduced without the combination. The fluctuations 
in the combined limits 
are due to the observed data events in the search for
first generation LQs. In the mass 
range from $250\,{\rm GeV}$ to $300\,{\rm GeV}$ both the combined limits on $\lambda_{\mu q}$ 
and $\lambda_{\tau q}$ are for all LQ types up to a factor 2 more stringent 
than without combination. Table \ref{masslimits} shows the 95\% CL combined lower 
limits on the LQ mass for all LQ types assuming a coupling of 
electromagnetic strength \mbox{$\lambda_{eq}\!=\!\lambda_{\mu q}(\lambda_{\tau q})\!=\!0.3$}.
\par
\begin{figure}
  \setlength{\unitlength}{1mm}
  \begin{picture}(138,155)
    \put(37,160){\Large{\bf H1~~~~Search for lepton flavour violation}}
    \put(0,80){\includegraphics[width=0.49\textwidth]{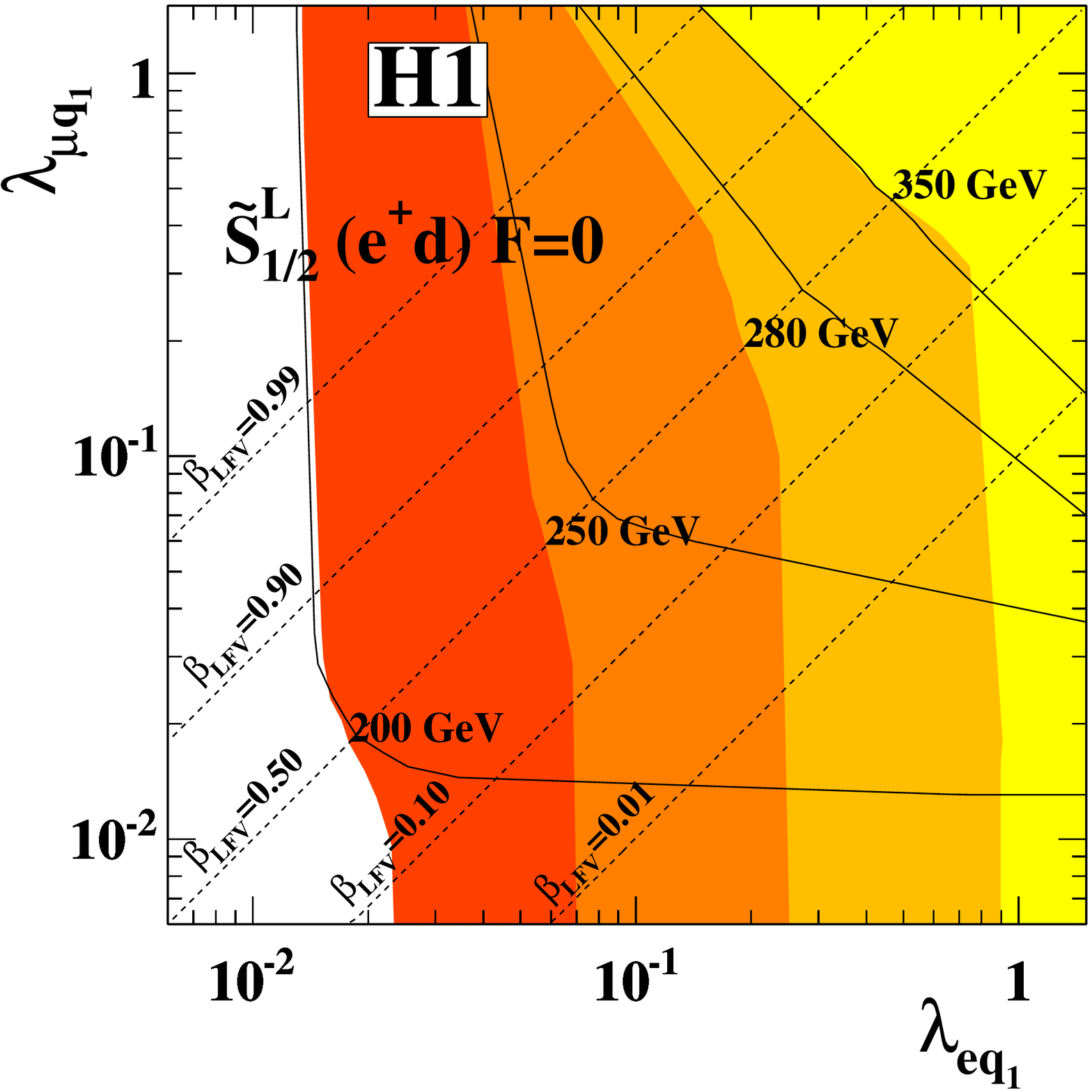}}
    \put(82,80){\includegraphics[width=0.49\textwidth]{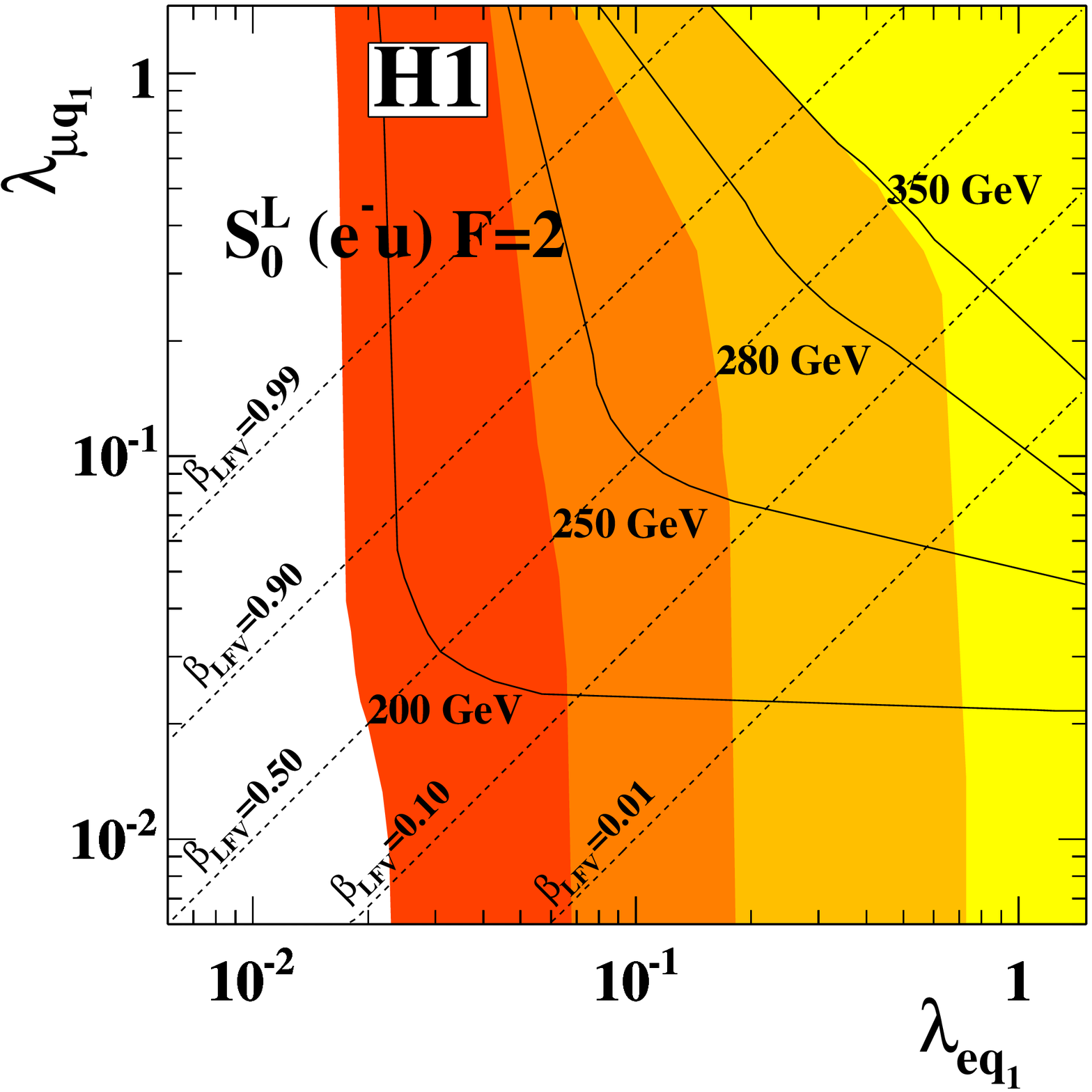}}
    \put(0,0){\includegraphics[width=0.49\textwidth]{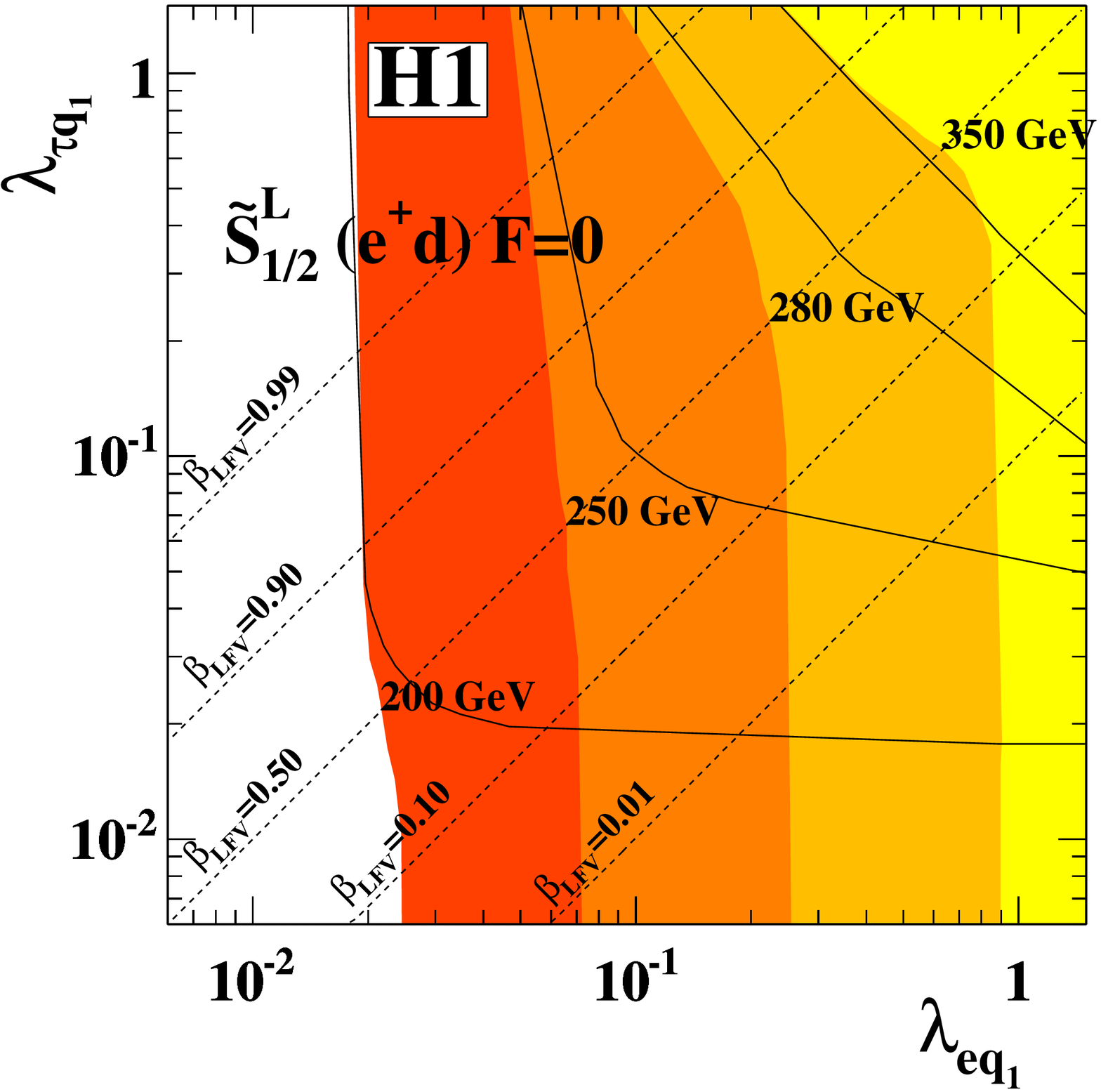}}
    \put(82,0){\includegraphics[width=0.49\textwidth]{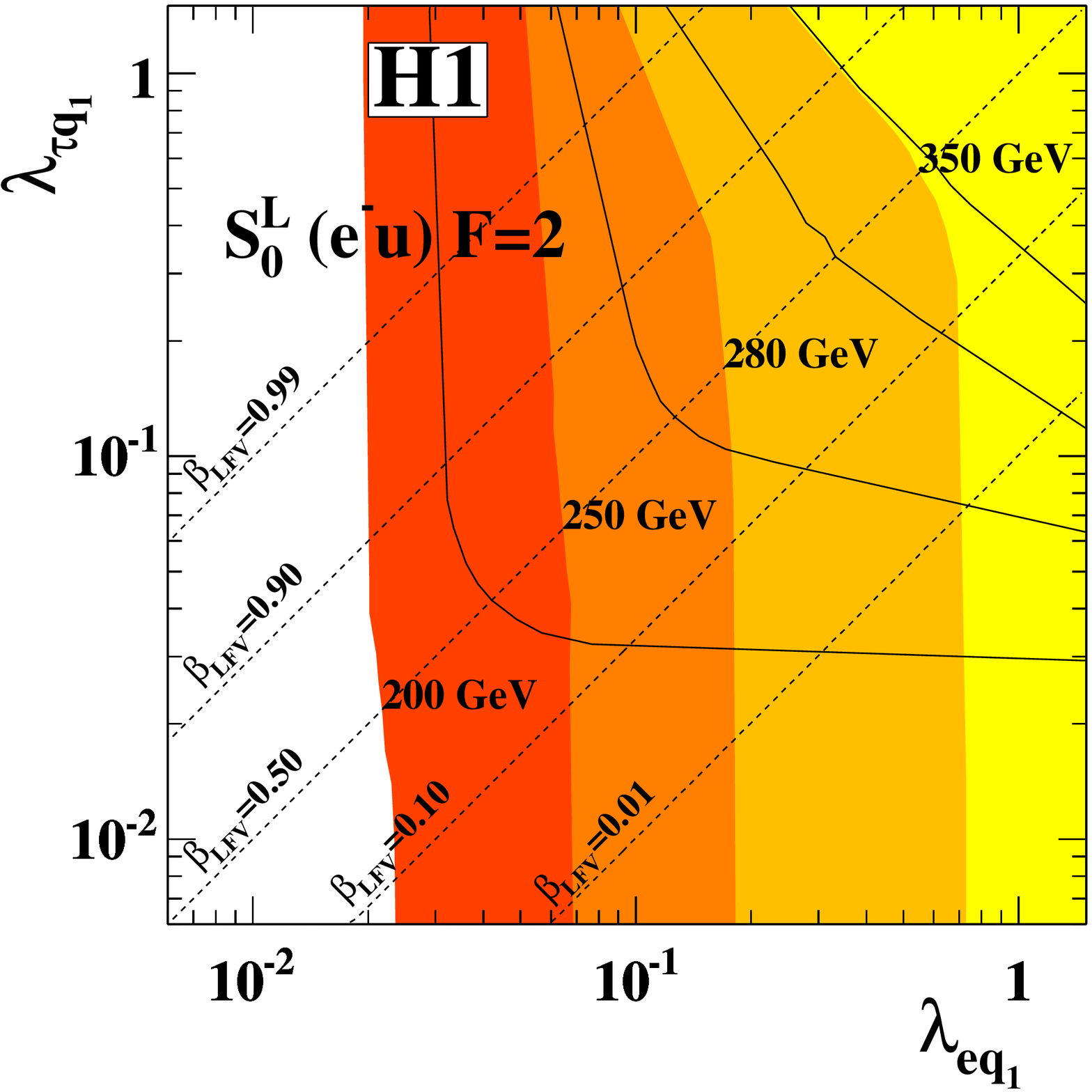}}
    \put(15,148){\small{\bf(a)}}
    \put(97,148){\small{\bf(b)}}
    \put(15,68){\small{\bf(c)}}
    \put(97,68){\small{\bf(d)}}
    \end{picture}
    \caption{
    Excluded regions at 95\% CL (filled) on $\lambda_{\ell q_1}$ as a function of
    $\lambda_{eq_1}$ for four different leptoquark masses. The branching ratio 
    \mbox{$\beta_{\rm LFV}=\lambda_{\ell q_1}^2/(\lambda_{\ell q_1}^2+\lambda_{eq_1}^2)$} is not
    fixed. Diagonal dashed lines represent iso-curves for fixed values of $\beta_{\rm
    LFV}$. The bounds deduced without the combination with first generation
    leptoquarks are shown as black curves corresponding to the different mass assumptions. 
    }
    \label{BRlim}
\end{figure}

Allowing for an arbitrary decay rate between the LFC and LFV decay channels, 
$\beta_{\rm  LFV}$, the excluded regions for two LQ types and four mass values in the
\mbox{$\lambda_{\mu q_1}\!-\!\lambda_{e q_1}$} (a,b) and \mbox{$\lambda_{\tau q_1}\!-\!\lambda_{e
  q_1}$} (c,d) planes are presented in figure \ref{BRlim}. 
For very low values of $\beta_{\rm LFV}$
(\mbox{$\lambda_{eq}\!\gg\!\lambda_{\mu q}(\lambda_{\tau q})$}), the limits
on $\lambda_{eq}$ reproduce the bounds published in \cite{h1lq}, as expected, since the LFC channel
dominates the LQ width. For \mbox{$\beta_{\rm LFV}\!\gg\!0.5$} 
(\mbox{$\lambda_{\mu q}(\lambda_{\tau q})\!\gg\!\lambda_{eq}$}) 
the present analysis extends significantly the published limits 
on $\lambda_{eq}$ to lower values. The limit without combination in the 
contact interaction region (where the cross section is proportional to 
\mbox{$\lambda_{eq_i}\lambda_{\mu(\tau)q_j}/m_{\rm  LQ}^2$}) 
forms a cross-diagonal straight line following different values of 
$\beta_{\rm LFV}$. The combination in the contact interaction
region, e.g. \mbox{$m_{\rm LQ}=350\,{\rm GeV}$}, 
barely strengthens the limit as the virtual effects of the 
high mass LQ contact interaction at low values of $\sqrt{\hat{s}}$ are marginal 
compared to the irreducible NC and CC DIS background. Fluctuations of 
the data may even result in a less stringent combined limit. 
\par
\begin{figure}
  \setlength{\unitlength}{1mm}
  \begin{picture}(138,155)
    \put(37,160){\Large{\bf H1~~~~Search for lepton flavour violation}}
    \put(0,80){\includegraphics[width=0.49\textwidth]{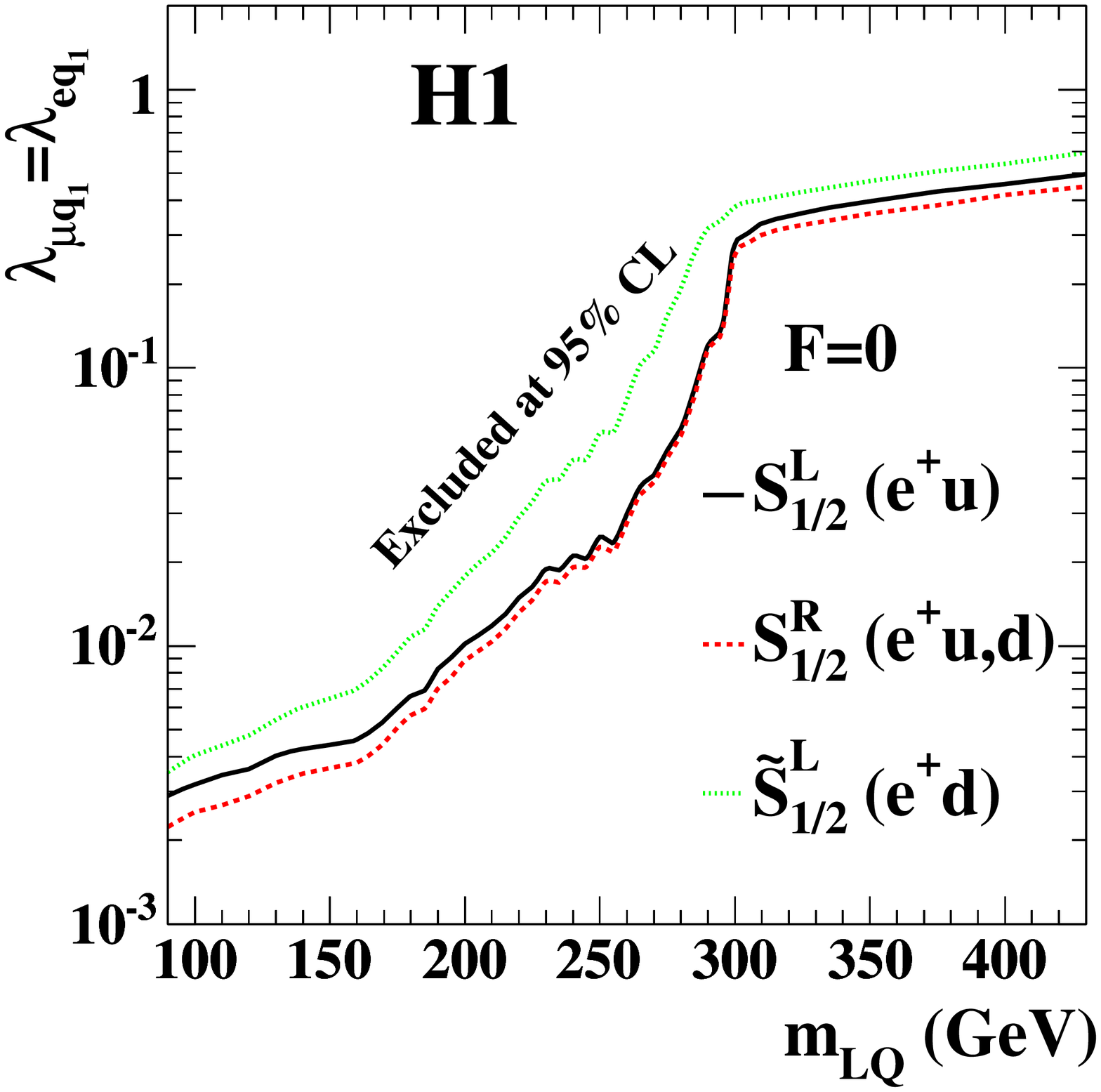}}
    \put(82,80){\includegraphics[width=0.49\textwidth]{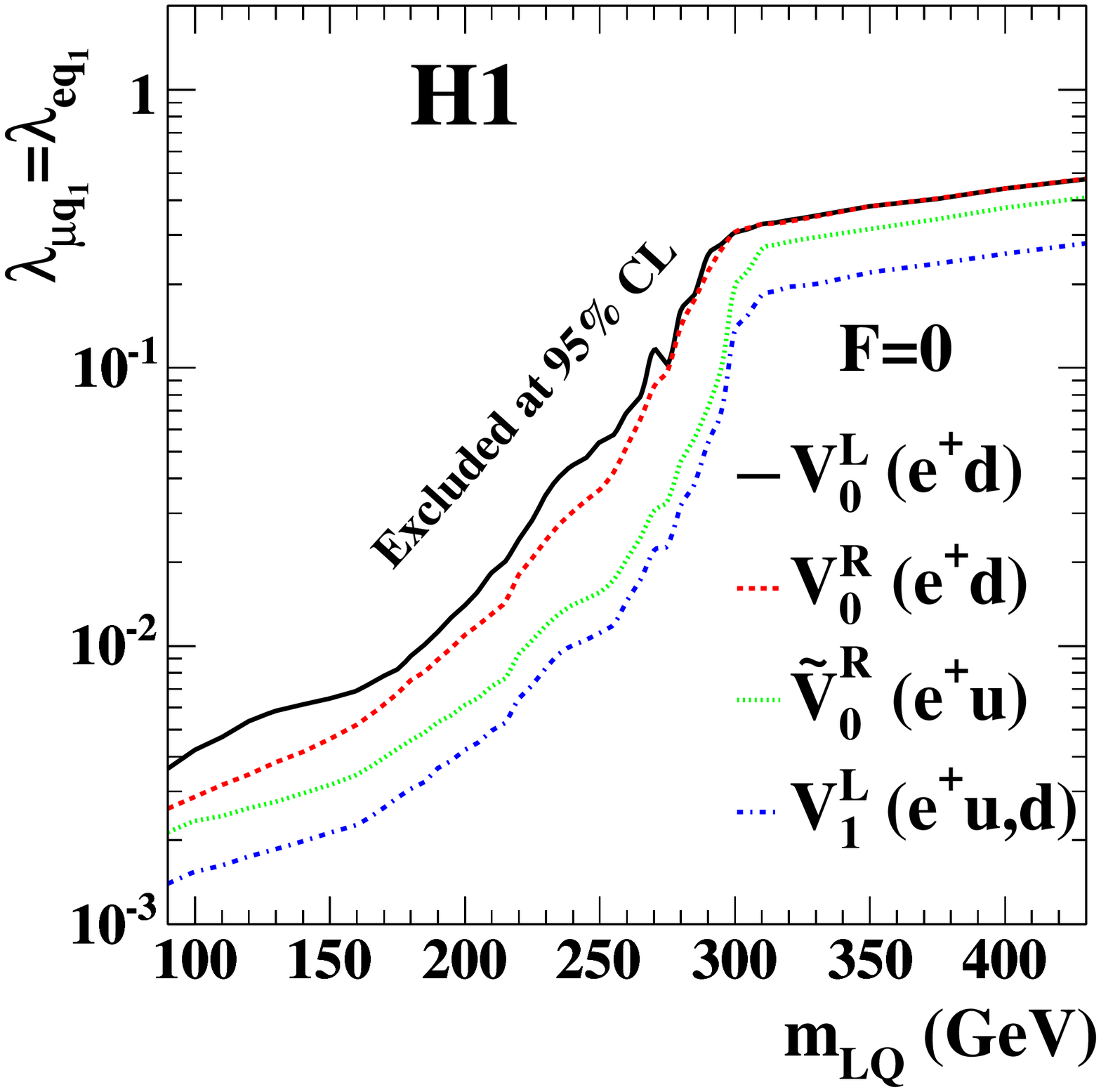}}
    \put(0,0){\includegraphics[width=0.49\textwidth]{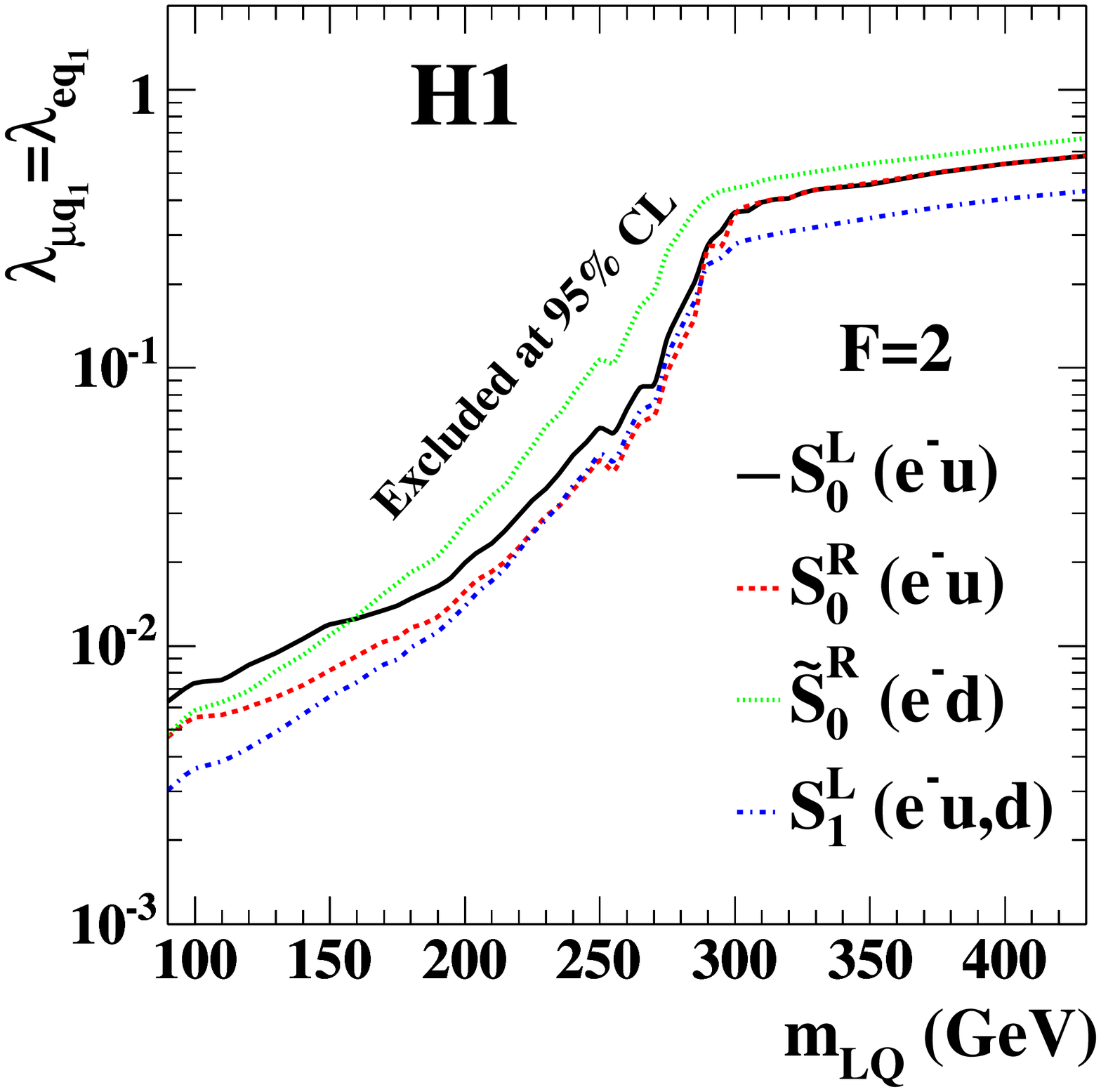}}
    \put(82,0){\includegraphics[width=0.49\textwidth]{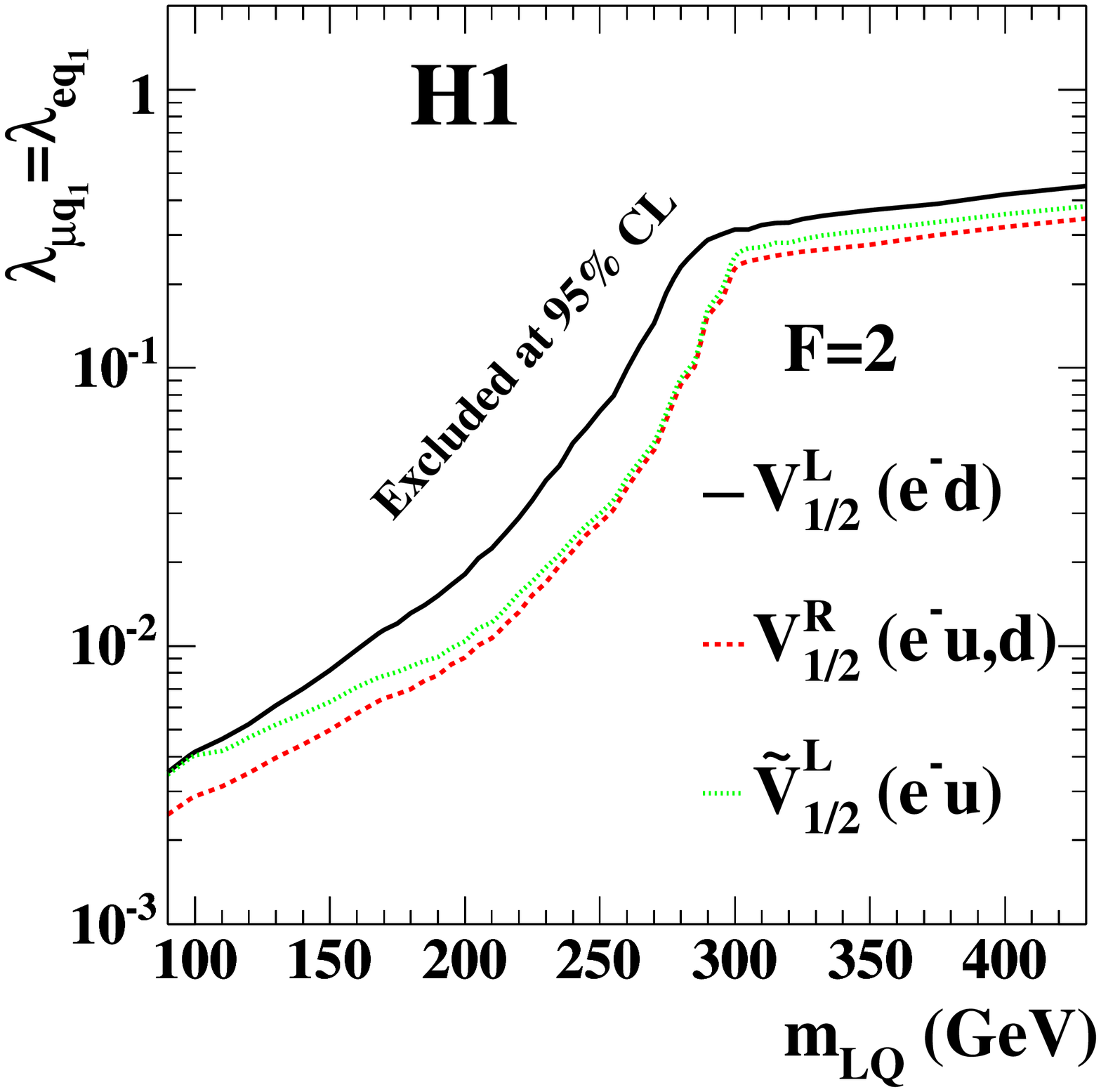}}
    \put(15,148){\small{\bf(a)}}
    \put(97,148){\small{\bf(b)}}
    \put(15,68){\small{\bf(c)}}
    \put(97,68){\small{\bf(d)}}
    \end{picture}
    \caption{
    Limits on the coupling constants $\lambda_{\mu q_1}\!=\!\lambda_{eq_1}$ 
    as a function of the leptoquark mass $m_{\rm LQ}$ 
    for  (a), (b) $F=0$ and  (c), (d) $F=2$ scalar and vector leptoquarks. 
    Regions above the lines are excluded at 95\% CL.
    The notation $q_1$ illustrates that only
    processes involving first generation quarks are considered. 
    }
    \label{figmuonlimits}
\end{figure}
\begin{figure}
  \setlength{\unitlength}{1mm}
  \begin{picture}(138,155)
    \put(37,160){\Large{\bf H1~~~~Search for lepton flavour violation}}
    \put(0,80){\includegraphics[width=0.49\textwidth]{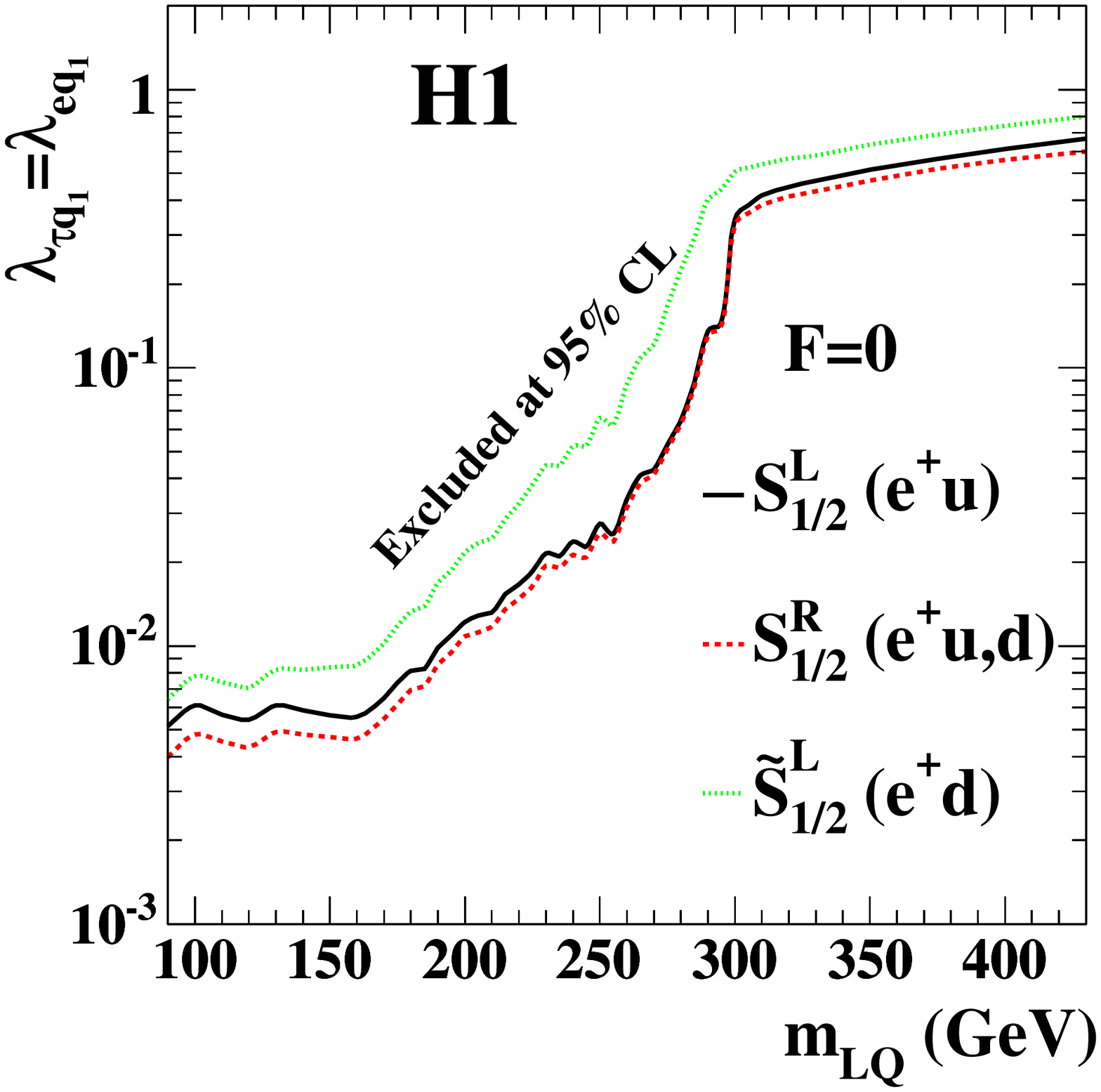}}
    \put(82,80){\includegraphics[width=0.49\textwidth]{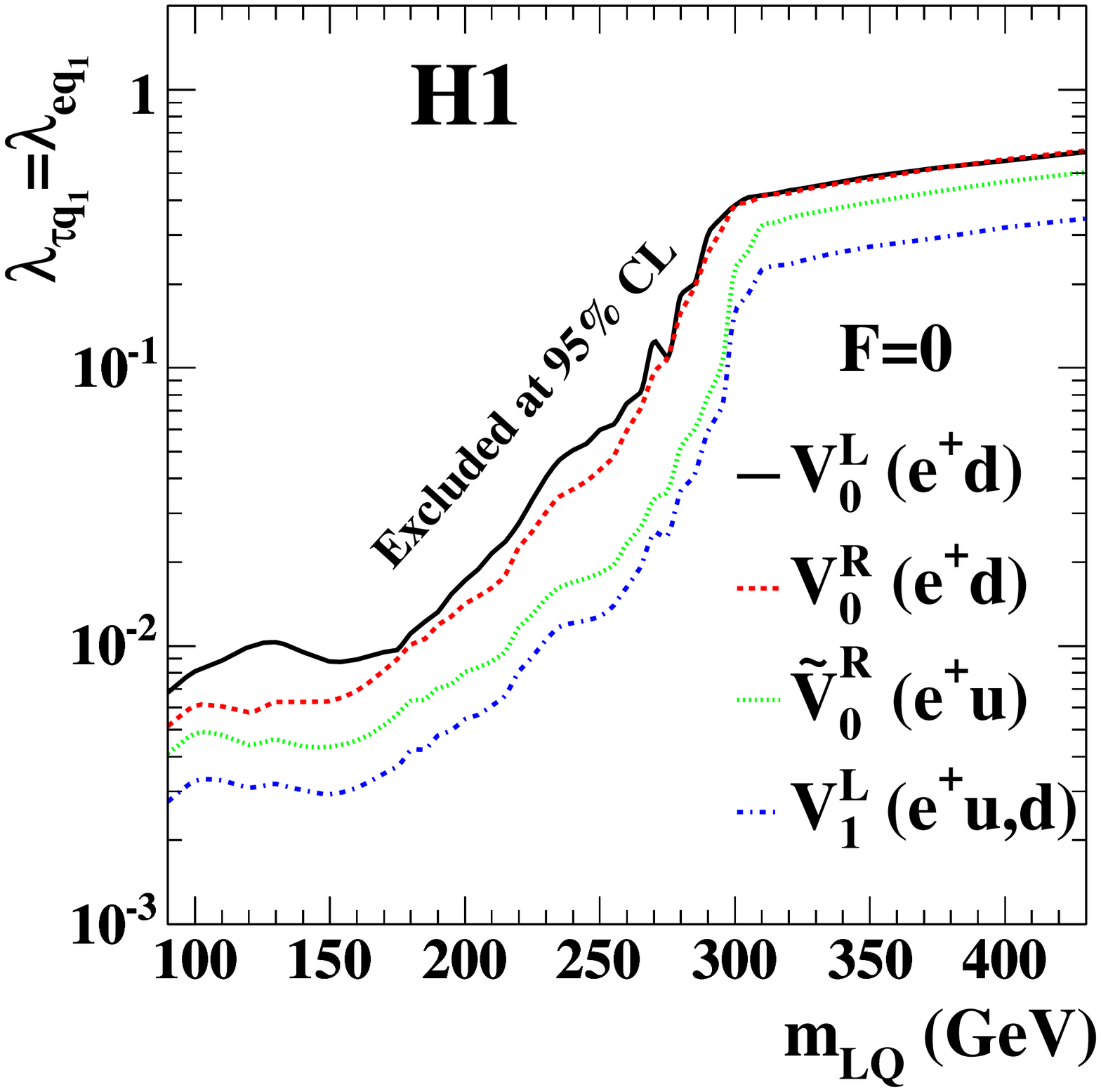}}
    \put(0,0){\includegraphics[width=0.49\textwidth]{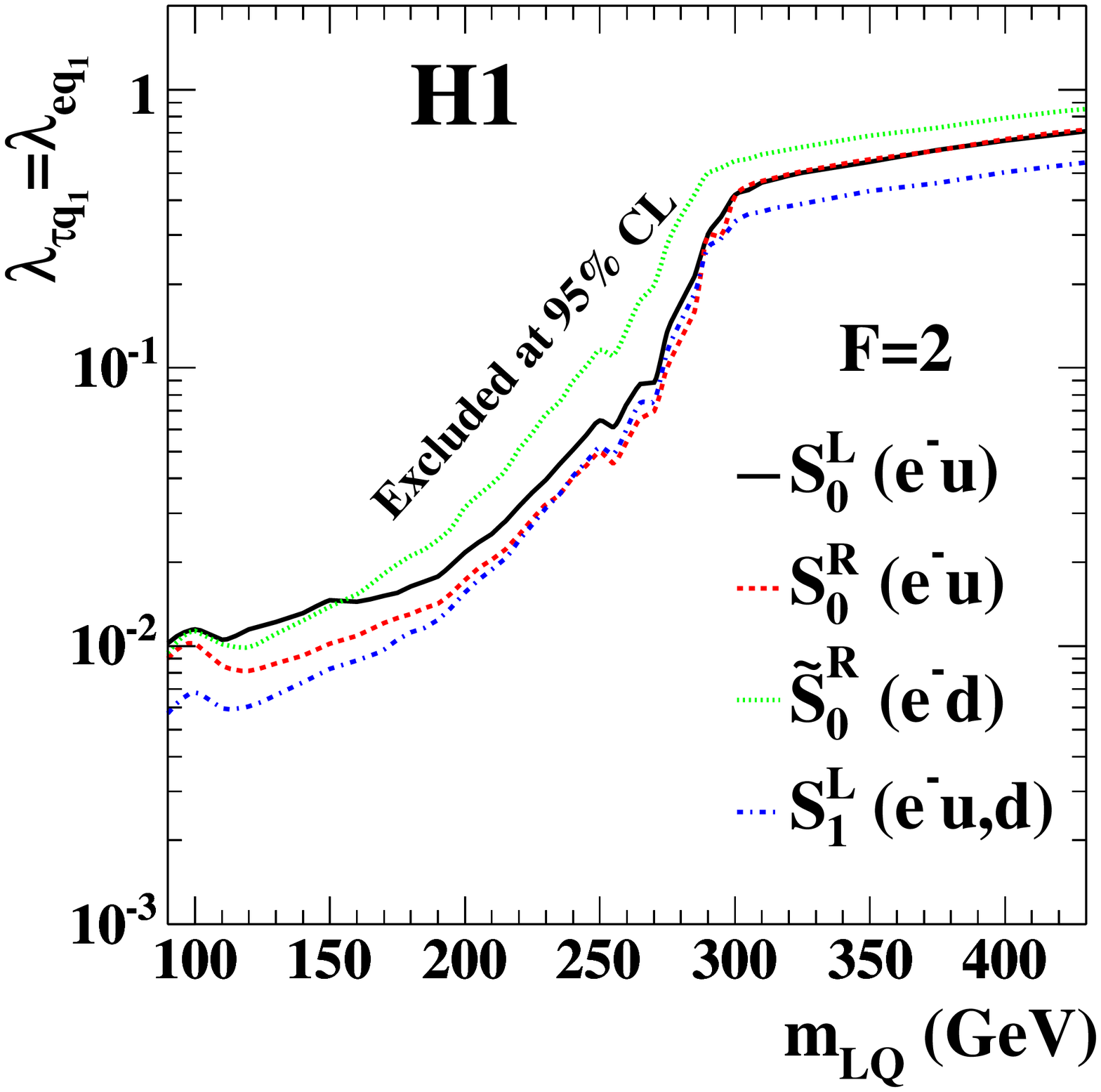}}
    \put(82,0){\includegraphics[width=0.49\textwidth]{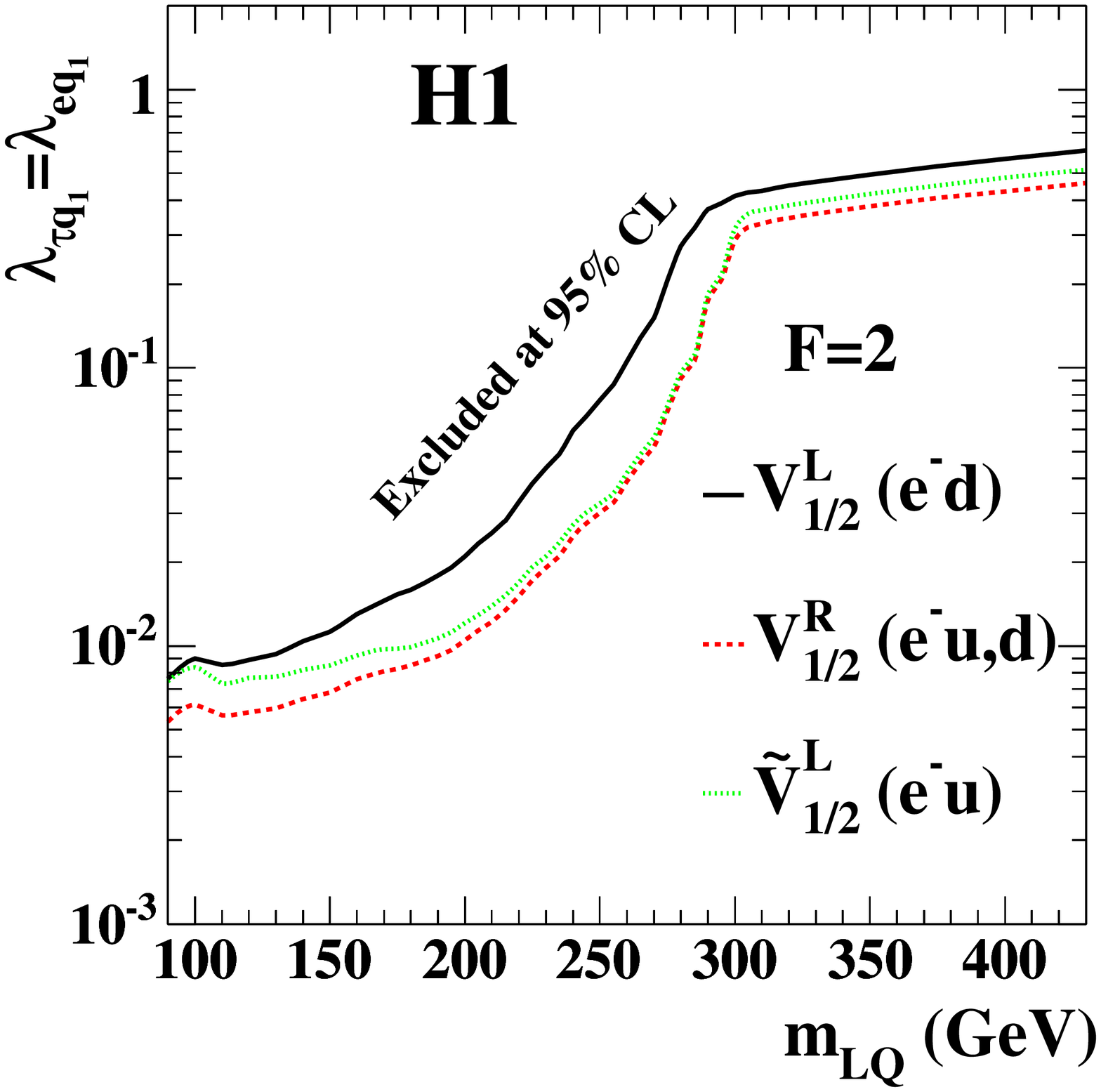}}
    \put(15,148){\small{\bf(a)}}
    \put(97,148){\small{\bf(b)}}
    \put(15,68){\small{\bf(c)}}
    \put(97,68){\small{\bf(d)}}
    \end{picture}
    \caption{
    Limits on the coupling constants $\lambda_{\tau q_1}\!=\!\lambda_{eq_1}$ 
    as a function of the leptoquark mass  $m_{\rm LQ}$
    for  (a), (b) $F=0$ and  (c), (d) $F=2$ scalar and vector leptoquarks. 
    Regions above the lines are excluded at 95\% CL.
    The notation $q_1$ illustrates that only
    processes involving first generation quarks are considered. 
    }
    \label{figtaulimits}
\end{figure}

Figures \ref{figmuonlimits} and \ref{figtaulimits} display the 95\% 
CL upper limits on the coupling 
$\lambda_{\mu q}$ and $\lambda_{\tau q}$ of all 14 LQ types to a muon-quark pair and a tau-quark pair, 
respectively, as a function of the LQ mass leading to LFV in $ep$
collisions, assuming \mbox{$\lambda_{eq}\!=\!\lambda_{\mu q}(\lambda_{\tau q})\!$}. The limit curves referring to the LQ types $S_0^L$ and
$\tilde{S}_{1/2}^L$ are identical to profiles of the corresponding excluded 
regions following the value \mbox{$\beta_{\rm LFV}\!=\!0.5$} in figure \ref{BRlim}. 
The limits are most stringent at low LQ masses with values $\mathcal{O}(10^{-3})$ around 
\mbox{$m_{\rm LQ}=100\,{\rm GeV}$}. Corresponding to the steeply falling parton density
function for high values of $x$, the LQ production cross section
decreases rapidly and exclusion limits are less stringent towards higher
LQ masses. For LQ mass values near the kinematical limit of 
$319\,{\rm GeV}$, the limit corresponding to a resonantly 
produced LQ turns smoothly into a limit on the virtual
effects of both an off-shell $s$-channel LQ process and a $u$-channel
LQ exchange. At masses $m_{\rm LQ}>\sqrt{s}$ the two processes contract to an effective 
four-fermion interaction, where the cross section is proportional to 
$(\lambda_{\mu(\tau)q}\lambda_{eq}/m^2_{\rm LQ})^2$. This feature is visible in the
constant increase of the exclusion limit for masses above the $ep$
centre-of-mass energy of $\sqrt{s}=319\,{\rm GeV}$. Due to initial state QED 
radiation and very low parton densities for masses near $\sqrt{s}$ the
``kink'' of the transition region is shifted to somewhat smaller masses of
around $290-300\,{\rm GeV}$.
\par
It is noticeable that the limits on vector
LQs are more stringent compared to those on the scalars, due to the 
considerably higher cross section and the slightly higher
acceptance. In each plot those LQ types that have couplings to
both $u$ and $d$ quarks exhibit the best limit. The limits corresponding to
LQs coupling to a $u$ quark are more stringent than those corresponding 
to LQs coupling to the $d$ quark only, as expected from the larger $u$ quark
density in the proton. The LQs $S_0^L$ and $S_0^R$ ($V_0^L$ and
$V_0^R$) differ only by the decay into a neutrino and a quark 
of the lefthanded LQ. As this decay channel is
not covered in the LFV decay channels, the left-handed LQ 
cannot be as strictly excluded as the right-handed one. This argument applies to the 
resonant production where the analysis is only sensitive to the partial width of
the LQ. In the high mass region the limits for $S_0^L$ and $S_0^R$ 
($V_0^L$ and $V_0^R$) are similar, as the four-fermion interaction is
independent of the decay width.
\par
The limits on \mbox{$\lambda_{\mu(\tau)q}\!=\!\lambda_{eq}$} derived from 
the virtual effects of a $500\,{\rm GeV}$ LQ are transformed into a limit on the value 
$\lambda_{\mu(\tau)q_j}\lambda_{eq_i}/m^2_{\rm LQ}$ and shown in tables 
\ref{F0muHighMassLQ} and \ref{F0tauHighMassLQ} for 
\mbox{$F=0$} LQs and in tables \ref{F2muHighMassLQ} and \ref{F2tauHighMassLQ} 
for \mbox{$F=2$} LQs. For each
LQ type the limit is calculated for the hypothesis of a process 
with only the quarks of flavours $i$ and $j$ involved. With respect to quark flavours, 
the selection criteria described in sections \ref{muon} and \ref{tau} 
are inclusive since no flavour tagging of the hadronic jet is used. 
\par
These results may be compared with constraints from low energy experiments, based
on the non-observation of LFV in muon scattering and rare decays of mesons and 
leptons~\cite{PDG}.
The interpretation in terms of leptoquark exchange and
limits on $\lambda_{\mu(\tau)q_j}\lambda_{eq_i}/m^2_{\rm LQ}$ \cite{DAVIDSON}
are also shown in tables \ref{F0muHighMassLQ}, \ref{F0tauHighMassLQ}, 
\ref{F2muHighMassLQ} and \ref{F2tauHighMassLQ}. 
Bounds of similar magnitude are observed for processes involving
$e\rightarrow\tau$ transitions and charm or bottom quarks. 
In these cases the limits obtained in the present analysis 
are often superior to those from low energy experiments.
\par
The results on LFV in LQ production are directly comparable with those from
the ZEUS experiment \cite{ZEUSLFV05}. Similar limits are obtained.
At hadron colliders LQs are mainly produced in pairs independently of the coupling,
and therefore searches cannot constrain LFV couplings. 
Lower mass limits on the second and third generation leptoquarks extend up to $250\,{\rm GeV}$  
and $150\,{\rm GeV}$, respectively, depending on the type and the assumed decay branching 
ratios \cite{TEVATRON}.
Similarly, second and third generation leptoquarks are pair produced in
$e^+e^-$ annihilation where typical lower mass bounds reach values of $100\,{\rm GeV}$~\cite{OPAL}.

\section{Conclusion}
\label{con}

A search for lepton flavour violation processes induced by leptoquarks
in $ep$ collisions at a centre-of-mass energy of $319\,{\rm GeV}$ 
with the H1 experiment at HERA is presented. No signal for 
the LFV processes \mbox{$ep\rightarrow\mu X$} or \mbox{$ep\rightarrow\tau X$} 
is found. Constraints on LFV LQ couplings are set combining the LFV
search with the search for first generation LQs. The limits are a factor 
of 2 to 4 more stringent and extend beyond the domain in LQ mass excluded 
by previous searches performed by the H1 experiment \cite{H1OLDLFV}. 
Exclusion limits on 
several scenarios of LFV transitions of the kind $eq_i\!\rightarrow\!\tau q_j$ 
are more stringent than limits from searches for certain rare meson or tau decays. 
Assuming a coupling of electromagnetic strength, leptoquarks mediating lepton flavour 
violating processes $e\!\rightarrow\!\mu$  and $e\!\rightarrow\!\tau$ can be ruled 
out up to masses of $459\,{\rm GeV}$ and $379\,{\rm GeV}$, respectively.

\section*{Acknowledgements}

We are grateful to the HERA machine group whose outstanding
efforts have made this experiment possible. 
We thank
the engineers and technicians for their work in constructing and
maintaining the H1 detector, our funding agencies for 
financial support, the
DESY technical staff for continual assistance
and the DESY directorate for support and for the
hospitality which they extend to the non DESY 
members of the collaboration.


\clearpage

\begin{table}
  \begin{small}
  \begin{center}
    \setlength{\extrarowheight}{3pt}
      \begin{tabular}{|c|c|c|c|rll|c|c|}
        \hline
         & & & & \multicolumn{3}{|c|}{\rule[-1.8mm]{0mm}{4mm}} & & Branching \\
        \up{Type} & \up{$J$} & \up{$F$} & \up{$Q$} &
        \multicolumn{3}{|c|}{\rule[-1.8mm]{0mm}{4mm} \up{$ep$ dominant process}} &
        \up{Coupling} & ratio $\beta_\ell$ \\
        \hdick
          & & & & & & $\ell^{-}u$ & $\lambda_L$ & $1/2$  \\
        \up{$S^L_0$} & \up{$0$} & \up{$2$} & \up{$-1/3$} & \up{$e^{-}_Lu_L$} &
        \up{$\rightarrow \Bigg\{ $} & $\nu_\ell d$ & $-\lambda_L$ & $1/2$  \\
        \hline
        $S^R_0$ & $0$ & $2$ & $-1/3$ & $e^{-}_Ru_R$ & $\rightarrow$ & $\ell^{-}u$
        & $\lambda_R$ & $1$ \\     
        \hline
        $\tilde{S}^R_0$ & $0$ & $2$ & $-4/3$ & $e^{-}_Rd_R$ & $\rightarrow$ & $\ell^{-}d$
        & $\lambda_R$ & $1$ \\
        \hline
        & & & & & & $\ell^{-}u$ & $-\lambda_L$ & $1/2$  \\
        $S^L_1$ & $0$ & $2$ & \up{$-1/3$} & \up{$e^{-}_Lu_L$} &
        \up{$\rightarrow \Bigg\{ $} & $\nu_\ell d$
        & $-\lambda_L$ & $1/2$ \\ 
        & & & $-4/3$ & $e^{-}_Ld_L$ & $\rightarrow$ & $\ell^{-}d$ &
        $-\sqrt{2}\lambda_L$ & $1$  \\
        \hline
        \hline
        $V^L_{1/2}$ & $1$ & $2$ & $-4/3$ & $e^{-}_Ld_R$ & $\rightarrow$ & $\ell^{-}d$
        & $\lambda_L$ & $1$ \\
        \hline
        & & & $-1/3$ & $e^{-}_Ru_L$ & $\rightarrow$ & $\ell^{-}u$ & $\lambda_R$ & $1$  \\
        \up{$V^R_{1/2}$} & \up{$1$} & \up{$2$} & $-4/3$ & $e^{-}_Rd_L$ &
        $\rightarrow$ & $\ell^{-}d$ & $\lambda_R$ & $1$  \\
        \hline
        $\tilde{V}^L_{1/2}$ & $1$ & $2$ & $-1/3$ & $e^{-}_Lu_R$ & $\rightarrow$ &
        $\ell^{-}u$ & $\lambda_L$ & $1$ \\
        \hdick
          & & & & & & $\ell^{+}d$ & $\lambda_L$ & $1/2$  \\
        \up{$V^L_0$} & \up{$1$} & \up{$0$} & \up{$+2/3$} & \up{$e^{+}_Rd_L$} &
        \up{$\rightarrow \Bigg\{ $} & $\bar{\nu}_\ell u$ & $\lambda_L$ & $1/2$  \\
        \hline
        $V^R_0$ & $1$ & $0$ & $+2/3$ & $e^{+}_Ld_R$ & $\rightarrow$ & $\ell^{+}d$
        & $\lambda_R$ & $1$ \\     
        \hline
        $\tilde{V}^R_0$ & $1$ & $0$ & $+5/3$ & $e^{+}_Lu_R$ & $\rightarrow$ & $\ell^{+}u$
        & $\lambda_R$ & $1$ \\
        \hline
        & & & & & & $\ell^{+}d$ & $-\lambda_L$ & $1/2$  \\
        $V^L_1$ & $1$ & $0$ & \up{$+2/3$} & \up{$e^{+}_Rd_L$} &
        \up{$\rightarrow \Bigg\{ $} & $\bar{\nu}_\ell u$
        & $\lambda_L$ & $1/2$ \\ 
        & & & $+5/3$ & $e^{+}_Ru_L$ & $\rightarrow$ & $\ell^{+}u$ &
        $\sqrt{2}\lambda_L$ & $1$  \\
        \hline
        \hline
        $S^L_{1/2}$ & $0$ & $0$ & $+5/3$ & $e^{+}_Ru_R$ & $\rightarrow$ & $\ell^{+}u$
        & $\lambda_L$ & $1$ \\
        \hline
        & & & $+2/3$ & $e^{+}_Ld_L$ & $\rightarrow$ & $\ell^{+}d$ & $-\lambda_R$ & $1$  \\
        \up{$S^R_{1/2}$} & \up{$0$} & \up{$0$} & $+5/3$ & $e^{+}_Lu_L$ &
        $\rightarrow$ & $\ell^{+}u$ & $\lambda_R$ & $1$  \\
        \hline
        $\tilde{S}^L_{1/2}$ & $0$ & $0$ & $+2/3$ & $e^{+}_Rd_R$ & $\rightarrow$ &
        $\ell^{+}d$ & $\lambda_L$ & $1$ \\
        \hline
      \end{tabular}
  \end{center}
  \end{small}
\caption[LQs in the Buchm\"uller-R\"uckl-Wyler classification.]
  {
    The 14 leptoquark (LQ) types of the Buchm\"uller-R\"uckl-Wyler
    classification~\cite{BRW} in the Aachen notation \cite{lego}. The LQ subscripts 
    refer to the weak isospin and the superscripts refer to the lepton
    chirality. Columns 2-4 display the spin $J$, fermion number $F$ and electrical
    charge $Q$. The dominant resonant production process in $ep$ scattering
    and the corresponding coupling is shown in columns 5 and 6
    respectively. Leptoquarks which couple to a left-handed lepton doublet and can
    decay into a neutrino-quark pair, have a charged lepton decay branching ratio of 
    $\beta_\ell\!=\!\Gamma_{\ell q}/(\Gamma_{\ell q}+\Gamma_{\nu_\ell q})\!=\!1/2$. 
  }
  \label{lqBRW}
\end{table}

\begin{table}
  \begin{center}
    \setlength{\extrarowheight}{3pt}
    \begin{tabular}{|c|c||c|c||c|c|c|c|c|}     
      \hline
      \multicolumn{1}{|c}{\rule[-2mm]{0mm}{5mm}{\bf H1}} &
      \multicolumn{3}{c}{\rule[-2mm]{0mm}{5mm} Search for LFV} &
      \multicolumn{5}{c|}{\rule[-2mm]{0mm}{5mm} $e^-p:~13.7\,{\rm pb}^{-1}~~~~~~$
           $e^+p:~66.5\,{\rm pb}^{-1}$} \\
         \hline
      \multicolumn{2}{|c||}{\rule[-2mm]{0mm}{5mm}} &
      \multicolumn{2}{c||}{\rule[-2mm]{0mm}{5mm} Selection results} &
      \multicolumn{5}{c|}{\rule[-2mm]{0mm}{5mm} Selection efficiency} \\
      \hdick
      \multicolumn{2}{|c||}{\rule[-2mm]{0mm}{5mm} Channel} & Data & SM MC & $m_{\rm LQ}$ & $S^R_0$ & $V^L_{1/2}$ & $V^R_0$ & $\tilde{S}^L_{1/2}$ \\
      \hline
      \hline
      & & & & $150\,{\rm GeV}$ & $58.0\%$ & $60.9\%$ & $60.1\%$ & $57.7\%$ \\
      & \up{$e^-p$} & \up{$0$} & \up{$0.18\pm0.06$} & $500\,{\rm GeV}$ & $47.2\%$ & $38.5\%$ & $42.3\%$ & $37.8\%$
      \\
      \cline{2-9}
      \up{$ep\rightarrow\mu X$} & & & & $150\,{\rm GeV}$ & $55.5\%$ & $57.9\%$ & $58.7\%$ & $55.8\%$ \\
      & \up{$e^+p$} & \up{$0$} & \up{$1.03\pm0.32$} & $500\,{\rm GeV}$ & $40.9\%$ & $40.5\%$ & $36.6\%$ & $41.4\%$
      \\
      \hline
      \hline
      & & & & $150\,{\rm GeV}$ & $28.3\%$ & $27.6\%$ & $27.1\%$ & $28.1\%$ \\
      & \up{$e^-p$} & \up{$0$} & \up{$0.75\pm0.21$} & $500\,{\rm GeV}$ & $21.3\%$ & $14.4\%$ & $17.1\%$ & $13.8\%$
      \\
      \cline{2-9}
      \up{$ep\rightarrow\tau X$} & & & & $150\,{\rm GeV}$ & $26.8\%$ & $26.4\%$ & $26.9\%$ & $27.0\%$ \\
      & \up{$e^+p$} & \up{$1$} & \up{$4.90\pm0.85$} & $500\,{\rm GeV}$ & $17.0\%$ & $16.7\%$ & $14.1\%$ & $17.3\%$
      \\
      \hline
      \hline
      \hline
      & & & & $150\,{\rm GeV}$ & $9.0\%$ & $7.8\%$ & $7.6\%$ & $8.9\%$ \\
      $ep\rightarrow\tau X$ & \up{$e^-p$} & \up{$0$} & \up{$0.28\pm0.19$} &
      $500\,{\rm GeV}$ & $6.7\%$ & $4.0\%$ & $5.2\%$ & $3.8\%$\\
       \cline{2-9}
      $\hookrightarrow \tau\rightarrow e\nu_e\nu_\tau$ & & & & $150\,{\rm GeV}$ & $8.3\%$ & $7.2\%$ & $7.3\%$ & $8.4\%$ \\
       & \up{$e^+p$} & \up{$0$} & \up{$1.24\pm0.55$} & $500\,{\rm GeV}$ & $4.8\%$ & $5.1\%$ & $4.0\%$ & $5.3\%$\\
      \hline
      \hline
      & & & & $150\,{\rm GeV}$ & $7.4\%$ & $7.6\%$ & $7.6\%$ & $7.4\%$ \\
      $ep\rightarrow\tau X$ & \up{$e^-p$} & \up{$0$} & \up{$0.18\pm0.06$} & $500\,{\rm GeV}$ & $6.3\%$ & $4.7\%$ & $5.4\%$ & $4.6\%$
      \\
       \cline{2-9}
      $\hookrightarrow \tau\rightarrow \mu\nu_\mu\nu_\tau$ & & & & $150\,{\rm GeV}$ & $7.8\%$ & $8.0\%$ & $8.1\%$ & $7.8\%$ \\
      & \up{$e^+p$} & \up{$0$} & \up{$1.03\pm0.32$} & $500\,{\rm GeV}$ & $5.2\%$ & $5.2\%$ & $4.5\%$ & $5.3\%$
      \\
      \hline
      \hline
      &  & & & $150\,{\rm GeV}$ & $11.9\%$ & $12.2\%$ & $11.9\%$ & $11.8\%$ \\
      $ep\rightarrow\tau X$& \up{$e^-p$} & \up{$0$} & \up{$0.29\pm0.06$} & $500\,{\rm GeV}$ & $8.3\%$ & $5.7\%$ & $6.5\%$ & $5.4\%$
      \\
       \cline{2-9}
      $\hookrightarrow\tau\rightarrow h\nu_\tau$ & & & & $150\,{\rm GeV}$ & $10.7\%$ & $11.2\%$ & $11.5\%$ & $10.8\%$ \\
      & \up{$e^+p$} & \up{$1$} & \up{$2.63\pm0.57$} & $500\,{\rm GeV}$ & $7.0\%$ & $6.4\%$ & $5.6\%$ & $6.7\%$
      \\
      \hline
    \end{tabular}
  \end{center}
  \caption[Summary of the selection results.]
  {Summary of the selection results of the search for the LFV processes
    $ep\rightarrow\mu X$ and $ep\rightarrow\tau X$. The results of the
    individual tau decay channels are also shown. The errors on the SM MC expectation include 
    statistical and systematic errors added in quadrature. Examples of 
    signal selection efficiencies for leptoquarks of the types $S^R_0$, $V^L_{1/2}$, $V^R_0$ and $S^L_{1/2}$ 
    with masses $m_{\rm LQ}$ of $150\,{\rm GeV}$ and $500\,{\rm GeV}$ are also
    shown. For the tau decay channels the efficiencies are normalised to the
    sum of all tau decays.}
  \label{summarytable}
\end{table}

\begin{table}
  \begin{center}
    \setlength{\extrarowheight}{5pt}
    \begin{tabular}{|c||c|c|c|c|c|c|c|}      
      \hline
      \multicolumn{8}{|c| }{\rule[-1.8mm]{0mm}{7mm}{\bf H1}~~~~lower 
        exclusion limits on $m_{\rm LQ}$ (GeV) at $95\%$ CL} 
          \\
      \hdick
     $F=0$ &    $S_{1/2}^{L}$  &
 $S_{1/2}^{R}$  & $\tilde{S}_{1/2}^{L}$ &
 $V_{0}^{L}$  & $V_{0}^{R}$   & $\tilde{V}_{0}^{R}$ & $V_{1}^{L}$ \\
      \hline 
       $eq \rightarrow \mu q$ & 302 & 309 & 288  & 299 & 298 & 333 & 459 \\
       $eq \rightarrow \tau q$ & 298 & 298 & 285 & 290 & 293 & 307 & 379 \\
      \hline
      \hline
      $F=2$ &    $S_{0}^{L}$  &
 $S_{0}^{R}$  & $\tilde{S}_{0}^{R}$ &
 $S_{1}^{L}$  & $V_{1/2}^{L}$   & $V_{1/2}^{R}$ & $\tilde{V}_{1/2}^{L}$ \\
      \hline   
       $eq \rightarrow \mu q$ & 294 & 294 & 278 & 306 & 299 & 374 & 336 \\
       $eq \rightarrow \tau q$ & 293 & 294 & 276 & 295 & 282 & 302 & 297 \\
       \hline
   \end{tabular}
 \end{center}
 \caption{Lower exclusion limits at 95\% CL on leptoquark masses $m_{\rm LQ}$ assuming 
   \mbox{$\lambda_{\mu q}\!=\!\lambda_{eq}=0.3$} or 
   \mbox{$\lambda_{\tau q}\!=\!\lambda_{eq}=0.3$}.}  
 \label{masslimits}
\end{table}

\begin{table}
\begin{center}
\setlength{\extrarowheight}{3pt}
\begin{tiny}

\begin{tabular}{|c||c|c|c|c|c|c|c|} 

\hline
\multicolumn{2}{|c}{\rule[-1.8mm]{0mm}{7mm}\Large{$ep \rightarrow \mu X$}}
& \multicolumn{4}{c}{\rule[-1.8mm]{0mm}{7mm}\Large{{\bf H1}}}
& \multicolumn{2}{c|}{\rule[-1.8mm]{0mm}{7mm}\Large{$F=0$}} \\

\hdick
\multicolumn{8}{|c|}{\rule[-1.8mm]{0mm}{7mm}\large{Upper exclusion 
  limits on $\lambda_{eq_i}\lambda_{\mu q_j}/m_{\rm LQ}^2~({\rm TeV}^{-2})$}}\\
\multicolumn{8}{|c|}{\rule[-1.8mm]{0mm}{7mm}\large{for lepton flavour
    violating leptoquarks at $95\%$ CL}}\\
\hline
  & \rule[-1.0mm]{0mm}{7mm}\large{$S_{1/2}^{L}$}  &
 \large{$S_{1/2}^{R}$}  & \large{$\tilde{S}_{1/2}^{L}$} &
 \large{$V_{0}^{L}$}  & \large{$V_{0}^{R}$}   & \large{$\tilde{V}_{0}^{R}$} & \large{$V_{1}^{L}$} \\
\raisebox{1.8ex}[-1.8ex]{\small $q_iq_j$ }      & $e^{-}\bar{u}$ & $e^{-}(\bar{u}+\bar{d})$ &    $e^{-}\bar{d}$     & $e^{-}\bar{d}$ & $e^{-}\bar{d}$ & $e^{-}\bar{u}$ & $e^{-}(\sqrt{2}\bar{u}+\bar{d})$ \\
  & $e^{+}u$ & $e^{+}(u+d)$ &    $e^{+}d$     & $e^{+}d$ & $e^{+}d$ & $e^{+}u$ & $e^{+}(\sqrt{2}u+d)$ \\
\hline \hline
 & $\mu N \rightarrow eN$ & $\mu N \rightarrow eN$ & $\mu N \rightarrow eN$ & $\mu N \rightarrow eN$ & $\mu N \rightarrow eN$ & $\mu N \rightarrow eN$ & $\mu N \rightarrow eN$  \\
  \small{1~1}  & $5.2 \times 10^{-5}$ & $2.6 \times 10^{-5}$ & $5.2 \times 10^{-5}$ & $2.6 \times 10^{-5}$ & $2.6 \times 10^{-5}$ & $2.6 \times 10^{-5}$ & $0.8 \times 10^{-5}$ \\
        & \bf \normalsize{1.4} & \bf \normalsize{1.1} & \bf \normalsize{2.0} & \bf \normalsize{1.3} & \bf \normalsize{1.3} & \bf \normalsize{0.9} & \bf \normalsize{0.4} \\
\hline

 & $D \rightarrow \mu \bar{e}$ & $K \rightarrow \mu \bar{e}$ & $K \rightarrow \mu \bar{e}$ & $K \rightarrow \mu \bar{e}$ & $K \rightarrow \mu \bar{e}$ & $D \rightarrow \mu \bar{e}$ & $K \rightarrow \mu \bar{e}$  \\
 \small{1~2}   & $2.4$ & $2 \times 10^{-5}$ & $2 \times 10^{-5}$ & $1 \times 10^{-5}$ & $1 \times 10^{-5}$ & $1.2$ & $1 \times 10^{-5}$ \\
        & \bf \normalsize{1.4} \cellcolor{orange} & \bf \normalsize{1.2} & \bf
        \normalsize2.0{} & \bf \normalsize{1.5} & \bf \normalsize{1.5} & \bf
        \normalsize{1.1} \cellcolor{orange} & \bf \normalsize{0.5} \\
\hline

    & & $B \rightarrow \mu \bar{e}$ & $B \rightarrow \mu \bar{e}$ & $B \rightarrow \mu \bar{e}$ & $B \rightarrow \mu \bar{e}$ &  & $B \rightarrow \mu \bar{e}$   \\
     \small{1~3}   & {\bf \large{$\ast$}} & $0.4$ & $0.4$ & $0.2$ & $0.2$ & {\bf \large{$\ast$}} & $0.2$ \\
        & & \bf \normalsize{2.1} & \bf \normalsize{2.1} & \bf \normalsize{1.6} & \bf \normalsize{1.6} & & \bf \normalsize{1.6} \\


\hline
\hline
 & $D \rightarrow \mu \bar{e}$ & $K \rightarrow \mu \bar{e}$ & $K \rightarrow \mu \bar{e}$ & $K \rightarrow \mu \bar{e}$ & $K \rightarrow \mu \bar{e}$ & $D \rightarrow \mu \bar{e}$ & $K \rightarrow \mu \bar{e}$  \\
  \small{2~1}  & $2.4$ & $2 \times 10^{-5}$ & $2 \times 10^{-5}$ & $1 \times 10^{-5}$ & $1 \times 10^{-5}$ & $1.2$ & $1 \times 10^{-5}$ \\
        & \bf \normalsize{4.2} & \bf \normalsize{2.9} & \bf \normalsize{4.1} & \bf \normalsize{1.7} & \bf \normalsize{1.7} & \bf \normalsize{1.5} & \bf \normalsize{0.7} \\
\hline

 & $\mu N \rightarrow eN$ & $\mu N \rightarrow eN$ & $\mu N \rightarrow eN$ & $\mu N \rightarrow eN$ & $\mu N \rightarrow eN$ & $\mu N \rightarrow eN$ & $\mu N \rightarrow eN$  \\
  \small{2~2}  & $9.2 \times 10^{-4}$  & $1.3 \times 10^{-3}$ & $3 \times 10^{-3}$ & $1.5 \times 10^{-3}$ & $1.5 \times 10^{-3}$ & $4.6 \times 10^{-4}$ & $2.7 \times 10^{-4}$ \\
        & \bf \normalsize{6.0} & \bf \normalsize{3.7} & \bf \normalsize{4.8} & \bf \normalsize{2.5} & \bf \normalsize{2.5} & \bf \normalsize{3.1} & \bf \normalsize{1.3} \\
\hline

    & & $B \rightarrow \bar{\mu} eK$ & $B \rightarrow \bar{\mu} eK$ & $B \rightarrow \bar{\mu} eK$ & $B \rightarrow \bar{\mu} eK$ &  & $B \rightarrow \bar{\mu} eK$   \\
      \small{2~3}  & {\bf \large{$\ast$}} & $0.3$ & $0.3$ & $0.15$ & $0.15$ & {\bf \large{$\ast$}} & $0.15$ \\
        & & \bf \normalsize{5.2} & \bf \normalsize{5.2} & \bf \normalsize{3.5} & \bf \normalsize{3.5} & & \bf \normalsize{3.5} \\


\hline
\hline
 & & $B \rightarrow \mu \bar{e}$ & $B \rightarrow \mu \bar{e}$ & $V_{ub}$ & $B \rightarrow \mu \bar{e}$ &  & $V_{ub}$  \\
  \small{3~1}  & {\bf \large{$\ast$}} & $0.4$ & $0.4$ & $0.12$ & $0.2$ & {\bf
    \large{$\ast$}} & $0.12$ \\
        &  & \bf \normalsize{5.3} & \bf \normalsize{5.3} & \bf \normalsize{1.8} & \bf \normalsize{1.8} &  & \bf \normalsize{1.8} \\
\hline

 &  & $B \rightarrow \bar{\mu}eK$ & $B \rightarrow \bar{\mu}eK$ & $B \rightarrow \bar{\mu}eK$ & $B \rightarrow \bar{\mu}eK$ & & $B \rightarrow \bar{\mu}eK$  \\
 \small{3~2}   & {\bf \large{$\ast$}} & $0.3$ & $0.3$ & $0.15$ & $0.15$ & {\bf
   \large{$\ast$}} & $0.15$ \\
        &  & \bf \normalsize{7.0} & \bf \normalsize{7.0} & \bf \normalsize{2.8} & \bf \normalsize{2.8} &  & \bf \normalsize{2.8} \\
\hline

   & & $\mu N \rightarrow eN$ & $\mu N \rightarrow eN$ & $\mu N \rightarrow eN$ & $\mu N \rightarrow eN$ &  & $\mu N \rightarrow eN$   \\
      \small{3~3}  & {\bf \large{$\ast$}} & $1.3 \times 10^{-3}$ & $3 \times 10^{-3}$ & $1.5 \times 10^{-3}$ & $1.5 \times 10^{-3}$ & {\bf \large{$\ast$}} & $2.7 \times 10^{-4}$ \\
        & & \bf \normalsize{8.3} & \bf \normalsize{8.3} & \bf \normalsize{4.3} & \bf \normalsize{4.3} & & \bf \normalsize{4.3} \\ 
\hline


\end{tabular}
\end{tiny}
\end{center}
\caption[Limits on $\lambda_{eq_i}\lambda_{\mu
  q_j}/m_{\rm LQ}^2$ for high mass $F=0$ leptoquarks]{
  Limits at 95\% CL on $\lambda_{eq_i}\lambda_{\mu
  q_j}/m_{\rm LQ}^2$ for $F=0$ leptoquarks (bold).
  Combinations of $i$ and $j$ shown in the first column denote 
  the quark generation coupling to the electron and muon respectively. 
  In each cell the first two rows show the process providing the most stringent limit 
  from low energy experiments. 
  The cases marked with '$\ast$' refer to scenarios involving a top
  quark, not considered in the present analysis. Highlighted H1 limits are more stringent than those from the
  corresponding low energy experiment.}
\label{F0muHighMassLQ}
\end{table}


\begin{table}
\begin{center}
\setlength{\extrarowheight}{3pt}
\begin{tiny}

\begin{tabular}{|c||c|c|c|c|c|c|c|} 

\hline
\multicolumn{2}{|c}{\rule[-1.8mm]{0mm}{7mm}\Large{$ep \rightarrow \tau X$}}
& \multicolumn{4}{c}{\rule[-1.8mm]{0mm}{7mm}\Large{{\bf H1}}}
& \multicolumn{2}{c|}{\rule[-1.8mm]{0mm}{7mm}\Large{$F=0$}} \\

\hdick
\multicolumn{8}{|c|}{\rule[-1.8mm]{0mm}{7mm}\large{Upper exclusion 
  limits on $\lambda_{eq_i}\lambda_{\tau q_j}/m_{\rm LQ}^2~({\rm TeV}^{-2})$}}\\
\multicolumn{8}{|c|}{\rule[-1.8mm]{0mm}{7mm}\large{for lepton flavour
    violating leptoquarks at $95\%$ CL}}\\
\hline
  & \rule[-1.0mm]{0mm}{7mm}\large{$S_{1/2}^{L}$}  &
 \large{$S_{1/2}^{R}$}  & \large{$\tilde{S}_{1/2}^{L}$} &
 \large{$V_{0}^{L}$}  & \large{$V_{0}^{R}$}   & \large{$\tilde{V}_{0}^{R}$} & \large{$V_{1}^{L}$} \\
\raisebox{1.8ex}[-1.8ex]{\small $q_iq_j$ }      & $e^{-}\bar{u}$ & $e^{-}(\bar{u}+\bar{d})$ &    $e^{-}\bar{d}$     & $e^{-}\bar{d}$ & $e^{-}\bar{d}$ & $e^{-}\bar{u}$ & $e^{-}(\sqrt{2}\bar{u}+\bar{d})$ \\
  & $e^{+}u$ & $e^{+}(u+d)$ &    $e^{+}d$     & $e^{+}d$ & $e^{+}d$ & $e^{+}u$ & $e^{+}(\sqrt{2}u+d)$ \\
\hline \hline
 & $\tau \rightarrow \pi e$ & $\tau \rightarrow \pi e$ & $\tau
  \rightarrow \pi e$ & $\tau \rightarrow \pi e$ & $\tau \rightarrow
  \pi e$ & $\tau \rightarrow \pi e$ & $\tau \rightarrow \pi e$  \\
\small{1~1} & $0.4$ & $0.2$ & $0.4$ & $0.2$ & $0.2$ & $0.2$ & $0.06$ \\
        & \bf \normalsize{2.1} & \bf \normalsize{1.8} & \bf \normalsize{3.1} & \bf \normalsize{2.1} & \bf \normalsize{2.1} & \bf \normalsize{1.5} & \bf \normalsize{0.7} \\
\hline

 & & $\tau \rightarrow Ke$ & $K \rightarrow \pi \nu \bar{\nu}$ & $\tau
 \rightarrow Ke$ & $\tau \rightarrow Ke$ &  & $K \rightarrow \pi \nu
 \bar{\nu}$  \\
 \small{1~2} & & $6.3$ & $5.8 \times 10^{-4}$ & $3.2$ & $3.2$ &  & $1.5 \times 10^{-4}$ \\
        & \bf \normalsize{2.2} \cellcolor{orange} & \bf \normalsize{1.8}
        \cellcolor{orange} & \bf \normalsize{3.2} & \bf \normalsize{2.5}
        \cellcolor{orange} & \bf \normalsize{2.5} \cellcolor{orange} & \bf
        \normalsize{1.7} \cellcolor{orange} & \bf \normalsize{0.8} \\
\hline

 & & $B \rightarrow \tau \bar{e}$ & $B \rightarrow \tau \bar{e}$ & $B \rightarrow \tau \bar{e}$ & $B \rightarrow \tau \bar{e}$ &  & $B \rightarrow \tau \bar{e}$   \\
 \small{1~3} & {\bf \large{$\ast$}} & $0.3$ & $0.3$ & $0.13$ & $0.13$ & {\bf \large{$\ast$}} & $0.13$ \\
        & & \bf \normalsize{3.2} & \bf \normalsize{3.2} & \bf \normalsize{2.7} & \bf \normalsize{2.7} & & \bf \normalsize{2.7} \\


\hline
\hline
 & & $\tau \rightarrow Ke$ & $K \rightarrow \pi \nu \bar{\nu}$ & $\tau
 \rightarrow Ke$ & $\tau \rightarrow Ke$ &  & $K \rightarrow \pi \nu
 \bar{\nu}$  \\
 \small{2~1} & & $6.3$ & $5.8 \times 10^{-4}$ & $3.2$ & $3.2$ &  & $1.5 \times 10^{-4}$ \\
        & \bf \normalsize{6.7} \cellcolor{orange} & \bf \normalsize{4.8}
        \cellcolor{orange} & \bf \normalsize{6.9} & \bf \normalsize{2.8}
        \cellcolor{orange} & \bf \normalsize{2.9} \cellcolor{orange} & \bf
        \normalsize{2.3} \cellcolor{orange} & \bf \normalsize{1.1} \\
\hline

 & $\tau \rightarrow 3 e$ & $\tau \rightarrow 3 e$ & $\tau \rightarrow 3 e$ & $\tau \rightarrow 3 e$ & $\tau \rightarrow 3 e$ & $\tau \rightarrow 3 e$ & $\tau \rightarrow 3 e$  \\
 \small{2~2} & $5.0$ & $8.0$ & $17.0$ & $9.0$ & $9.0$ & $3.0$ & $1.6$ \\
        & \bf \normalsize{10.9} & \bf \normalsize{6.7} \cellcolor{orange} &
        \bf \normalsize{8.6} \cellcolor{orange} & \bf \normalsize{4.5}
        \cellcolor{orange} & \bf \normalsize{4.5} \cellcolor{orange} & \bf \normalsize{5.5} & \bf \normalsize{2.4} \\
\hline

    & & $B \rightarrow \tau \bar{e}X$ & $B \rightarrow \tau \bar{e}X$ & $B \rightarrow \tau \bar{e}X$ & $B \rightarrow \tau \bar{e}X$ &  & $B \rightarrow \tau \bar{e}X$   \\
      \small{2~3}  & {\bf \large{$\ast$}} & $14.0$ & $14.0$ & $7.2$ & $7.2$ & {\bf \large{$\ast$}} & $7.2$ \\
        & & \bf \normalsize{9.3} \cellcolor{orange} & \bf \normalsize{9.3}
        \cellcolor{orange} & \bf \normalsize{6.3} \cellcolor{orange} & \bf
        \normalsize{6.3} \cellcolor{orange} & & \bf \normalsize{6.3} \cellcolor{orange} \\


\hline
\hline
 & & $B \rightarrow \tau \bar{e}$ & $B \rightarrow \tau \bar{e}$ & $V_{ub}$ & $B \rightarrow \tau \bar{e}$ &  & $V_{ub}$  \\
    \small{3~1} & {\bf \large{$\ast$}} & $0.3$ & $0.3$ & $0.12$ & $0.13$ &
    {\bf \large{$\ast$}} & $0.12$ \\
        &  & \bf \normalsize{9.1} & \bf \normalsize{9.1} & \bf \normalsize{3.0} & \bf \normalsize{3.0} &  & \bf \normalsize{3.0} \\
\hline

 &  & $B \rightarrow \tau \bar{e}X$ & $B \rightarrow \tau \bar{e}X$ & $B \rightarrow \tau \bar{e}X$ & $B \rightarrow \tau \bar{e}X$ & & $B \rightarrow \tau \bar{e}X$  \\
   \small{3~2} & {\bf \large{$\ast$}} & $14.0$ & $14.0$ & $7.2$ & $7.2$ & {\bf
     \large{$\ast$}} & $7.2$ \\
        &  & \bf \normalsize{12.6} \cellcolor{orange} & \bf \normalsize{12.6}
        \cellcolor{orange} & \bf \normalsize{4.9} \cellcolor{orange} & \bf
        \normalsize{4.9} \cellcolor{orange} &  & \bf \normalsize{4.9}
        \cellcolor{orange} \\
\hline

   & & $\tau \rightarrow 3 e$ & $\tau \rightarrow 3 e$ & $\tau \rightarrow 3 e$ & $\tau \rightarrow 3 e$ &  & $\tau \rightarrow 3 e$   \\
       \small{3~3} & {\bf \large{$\ast$}} & $8.0$ & $17.0$ & $9.0$ & $9.0$ & {\bf \large{$\ast$}} & $1.6$ \\
        & & \bf \normalsize{15.2} & \bf \normalsize{15.2} \cellcolor{orange} &
        \bf \normalsize{8.1} \cellcolor{orange} & \bf \normalsize{8.1}
        \cellcolor{orange} & & \bf \normalsize{8.1} \\ 
\hline


\end{tabular}
\end{tiny}
\end{center}
\caption[Limits on $\lambda_{eq_i}\lambda_{\tau
  q_j}/m_{\rm LQ}^2$ for high mass $F=0$ LQs]{
  Limits at 95\% CL on $\lambda_{eq_i}\lambda_{\tau
  q_j}/m_{\rm LQ}^2$ for $F=0$ leptoquarks (bold).
  Combinations of $i$ and $j$ shown in the first column denote 
  the quark generation coupling to the electron and tau respectively. 
  In each cell the first two rows show the process providing the most stringent limit 
  from low energy experiments. 
  The cases marked with '$\ast$' refer to scenarios involving a top quark, not considered in the present analysis. 
  Highlighted H1 limits
  are more stringent than those from the corresponding low energy experiment.}
\label{F0tauHighMassLQ}
\end{table}

\begin{table}
\begin{center}
\setlength{\extrarowheight}{3pt}
\begin{tiny}

\begin{tabular}{|c||c|c|c|c|c|c|c|} 

\hline
\multicolumn{2}{|c}{\rule[-1.8mm]{0mm}{7mm}\Large{$ep \rightarrow \mu X$}}
& \multicolumn{4}{c}{\rule[-1.8mm]{0mm}{7mm}\Large{{\bf H1}}}
& \multicolumn{2}{c|}{\rule[-1.8mm]{0mm}{7mm}\Large{$F=2$}} \\

\hdick
\multicolumn{8}{|c|}{\rule[-1.8mm]{0mm}{7mm}\large{Upper exclusion 
  limits on $\lambda_{eq_i}\lambda_{\mu q_j}/m_{\rm LQ}^2~({\rm TeV}^{-2})$}}\\
\multicolumn{8}{|c|}{\rule[-1.8mm]{0mm}{7mm}\large{for lepton flavour
    violating leptoquarks at $95\%$ CL}}\\
\hline
  & \rule[-1.0mm]{0mm}{7mm}\large{$S_{0}^{L}$}  &
 \large{$S_{0}^{R}$}  & \large{$\tilde{S}_{0}^{R}$} &
 \large{$S_{1}^{L}$}  & \large{$V_{1/2}^{L}$}   & \large{$V_{1/2}^{R}$} & \large{$\tilde{V}_{1/2}^{L}$} \\
\raisebox{1.8ex}[-1.8ex]{\small $q_iq_j$ }      & $e^{-}u$ &
  $e^{-}u$ &    $e^{-}(u+d)$  & $e^{-}(u+\sqrt{2}d)$   & $e^{-}d$ &
  $e^{-}(u+d)$ & $e^{-}u$ \\
      & $e^{+}\bar{u}$ &
  $e^{+}\bar{u}$ &    $e^{+}(\bar{u}+\bar{d})$  & $e^{+}(\bar{u}+\sqrt{2}\bar{d})$   & $e^{+}\bar{d}$ &
  $e^{+}(\bar{u}+\bar{d})$ & $e^{+}\bar{u}$ \\
\hline \hline
 & $\mu N \rightarrow eN$ & $\mu N \rightarrow eN$ & $\mu N \rightarrow eN$ & $\mu N \rightarrow eN$ & $\mu N \rightarrow eN$ & $\mu N \rightarrow eN$ & $\mu N \rightarrow eN$  \\
  \small{1~1}  & $5.2 \times 10^{-5}$ & $5.2 \times 10^{-5}$ & $5.2 \times 10^{-5}$ & $1.7 \times 10^{-5}$ & $2.6 \times 10^{-5}$ & $1.3 \times 10^{-5}$ & $2.6 \times 10^{-5}$ \\
        & \bf \normalsize{2.0} & \bf \normalsize{2.0} & \bf \normalsize{2.6} & \bf \normalsize{1.0} & \bf \normalsize{1.1} & \bf \normalsize{0.6} & \bf \normalsize{0.8} \\
\hline

 & $K \rightarrow \pi \nu \bar{\nu}$ & $D \rightarrow \mu \bar{e}$ & $K \rightarrow \mu \bar{e}$ & $K \rightarrow \mu \bar{e}$ & $K \rightarrow \mu \bar{e}$ & $K \rightarrow \mu \bar{e}$ & $D \rightarrow \mu \bar{e}$  \\
   \small{1~2} & $1 \times 10^{-3}$ & $2.4$ & $2 \times 10^{-5}$ & $1 \times
    10^{-5}$ & $1 \times 10^{-5}$ & $1 \times 10^{-5}$ & $1.2$ \\
        & \bf \normalsize{2.6} & \bf \normalsize{2.6} & \bf \normalsize{3.2} & \bf \normalsize{1.4} & \bf \normalsize{2.0} & \bf \normalsize{1.4} & \bf \normalsize{1.9} \\
\hline

    & & & $B \rightarrow \mu \bar{e}$ & $V_{ub}$ & $B \rightarrow \mu \bar{e}$ & $B \rightarrow \mu \bar{e}$ &  \\
   \small{1~3} & {\bf \large{$\ast$}} & {\bf \large{$\ast$}} & $0.4$ & $0.24$
   & $0.2$ & $0.2$ & {\bf \large{$\ast$}} \\
    & & & \bf \normalsize{3.3} & \bf \normalsize{1.6} & \bf
    \normalsize{2.5} & \bf \normalsize{2.5} & \\


\hline
\hline
 & $K \rightarrow \pi \nu \bar{\nu}$ & $D \rightarrow \mu \bar{e}$ & $K \rightarrow \mu \bar{e}$ & $K \rightarrow \mu \bar{e}$ & $K \rightarrow \mu \bar{e}$ & $K \rightarrow \mu \bar{e}$ & $D \rightarrow \mu \bar{e}$  \\
  \small{2~1}  & $1 \times 10^{-3}$ & $2.4$ & $2 \times 10^{-5}$ & $1 \times 10^{-5}$ & $1 \times 10^{-5}$ & $1 \times 10^{-5}$ & $1.2$ \\
        & \bf \normalsize{2.6} & \bf \normalsize{2.6} & \bf \normalsize{3.3} & \bf \normalsize{1.4} & \bf \normalsize{1.1} & \bf \normalsize{0.7} & \bf \normalsize{0.8} \cellcolor{orange}\\
\hline

 & $\mu N \rightarrow eN$ & $\mu N \rightarrow eN$ & $\mu N \rightarrow eN$ & $\mu N \rightarrow eN$ & $\mu N \rightarrow eN$ & $\mu N \rightarrow eN$ & $\mu N \rightarrow eN$  \\
  \small{2~2}  & $9.2 \times 10^{-4}$  & $9.2 \times 10^{-3}$ & $3 \times 10^{-3}$ & $2.5 \times 10^{-3}$ & $1.5 \times 10^{-3}$ & $6.7 \times 10^{-4}$ & $4.6 \times 10^{-4}$ \\
        & \bf \normalsize{6.0} & \bf \normalsize{6.0} & \bf \normalsize{4.8} & \bf \normalsize{2.2} & \bf \normalsize{2.5} & \bf \normalsize{1.9} & \bf \normalsize{3.1} \\
\hline

    & & & $B \rightarrow \bar{\mu} eK$ & $B \rightarrow \bar{\mu} eK$ & $B \rightarrow \bar{\mu} eK$ & $B \rightarrow \bar{\mu} eK$ &  \\
    \small{2~3}  & {\bf \large{$\ast$}} & {\bf \large{$\ast$}} & $0.3$ & $0.15$ & $0.15$ & $0.15$ & {\bf \large{$\ast$}}\\
      &  & & \bf \normalsize{5.2} & \bf \normalsize{2.6} &
      \bf\normalsize{3.5} & \bf \normalsize{3.5} &
      \\


\hline
\hline
 & & & $B \rightarrow \mu \bar{e}$ & $B \rightarrow \mu \bar{e}$ & $B \rightarrow \mu \bar{e}$ & $B \rightarrow \mu \bar{e}$ &    \\
 \small{3~1} & {\bf \large{$\ast$}} & {\bf \large{$\ast$}} & $0.4$ & $0.4$ & $0.2$ & $0.2$ & {\bf \large{$\ast$}} \\
     
 & & & \bf \normalsize{3.7} & \bf \normalsize{1.9} & \bf \normalsize{1.2} & \bf \normalsize{1.2} & \\
\hline

 & & & $B \rightarrow \bar{\mu}eK$ & $B \rightarrow \bar{\mu}eK$ & $B \rightarrow \bar{\mu}eK$ & $B \rightarrow \bar{\mu}eK$ & \\
 \small{3~2} & {\bf \large{$\ast$}} & {\bf \large{$\ast$}} & $0.3$ & $0.15$ &
 $0.15$ & $0.15$ & {\bf \large{$\ast$}} \\
 & & & \bf \normalsize{7.0} & \bf \normalsize{3.5} & \bf \normalsize{2.8} & \bf \normalsize{2.8} &   \\
\hline

 & & & $\mu N \rightarrow eN$ & $\mu N \rightarrow eN$ & $\mu N \rightarrow eN$ & $\mu N \rightarrow eN$ &    \\
 \small{3~3} & {\bf \large{$\ast$}} & {\bf \large{$\ast$}} & $3 \times 10^{-3}$ & $2.5 \times 10^{-3}$ & $1.5 \times 10^{-3}$ & $6.7 \times 10^{-4}$ & {\bf \large{$\ast$}}  \\
 & & & \bf \normalsize{8.3} & \bf \normalsize{4.3} & \bf \normalsize{4.3} & \bf \normalsize{4.3} &  \\ 
\hline


\end{tabular}
\end{tiny}
\end{center}
\caption[Limits on $\lambda_{eq_i}\lambda_{\mu
  q_j}/m_{\rm LQ}^2$ for high mass $F=2$ LQs]{
  Limits at 95\% CL on $\lambda_{eq_i}\lambda_{\mu
  q_j}/m_{\rm LQ}^2$ for $F=2$ leptoquarks (bold).
  Combinations of $i$ and $j$ shown in the first column denote 
  the quark generation coupling to the electron and muon respectively. 
  In each cell the first two rows show the process providing the most stringent limit 
  from low energy experiments. 
  The cases marked with '$\ast$' refer to scenarios involving a top
  quark, not considered in the present analysis. Highlighted H1 limits are more stringent than those from the
  corresponding low energy experiment.}
\label{F2muHighMassLQ}
\end{table}

\begin{table}
\begin{center}
\setlength{\extrarowheight}{3pt}
\begin{tiny}

\begin{tabular}{|c||c|c|c|c|c|c|c|} 

\hline
\multicolumn{2}{|c}{\rule[-1.8mm]{0mm}{7mm}\Large{$ep \rightarrow \tau X$}}
& \multicolumn{4}{c}{\rule[-1.8mm]{0mm}{7mm}\Large{{\bf H1}}}
& \multicolumn{2}{c|}{\rule[-1.8mm]{0mm}{7mm}\Large{$F=2$}} \\

\hdick
\multicolumn{8}{|c|}{\rule[-1.8mm]{0mm}{7mm}\large{Upper exclusion 
  limits on $\lambda_{eq_i}\lambda_{\tau q_j}/m_{\rm LQ}^2~({\rm TeV}^{-2})$}}\\
\multicolumn{8}{|c|}{\rule[-1.8mm]{0mm}{7mm}\large{for lepton flavour
    violating leptoquarks at $95\%$ CL}}\\
\hline
  & \rule[-1.0mm]{0mm}{7mm}\large{$S_{0}^{L}$}  &
 \large{$S_{0}^{R}$}  & \large{$\tilde{S}_{0}^{R}$} &
 \large{$S_{1}^{L}$}  & \large{$V_{1/2}^{L}$}   & \large{$V_{1/2}^{R}$} & \large{$\tilde{V}_{1/2}^{L}$} \\
\raisebox{1.8ex}[-1.8ex]{\small $q_iq_j$ }      & $e^{-}u$ &
  $e^{-}u$ &    $e^{-}(u+d)$  & $e^{-}(u+\sqrt{2}d)$   & $e^{-}d$ &
  $e^{-}(u+d)$ & $e^{-}u$ \\
      & $e^{+}\bar{u}$ &
  $e^{+}\bar{u}$ &    $e^{+}(\bar{u}+\bar{d})$  & $e^{+}(\bar{u}+\sqrt{2}\bar{d})$   & $e^{+}\bar{d}$ &
  $e^{+}(\bar{u}+\bar{d})$ & $e^{+}\bar{u}$ \\
\hline \hline
 & $G_F$ & $\tau \rightarrow \pi e$ & $\tau \rightarrow \pi e$ & $\tau \rightarrow \pi e$ & $\tau \rightarrow \pi e$ & $\tau \rightarrow \pi e$ & $\tau \rightarrow \pi e$  \\
   \small{1~1} & $0.3$ & $0.4$ & $0.4$ & $0.1$ & $0.2$ & $0.1$ & $0.2$ \\
        & \bf \normalsize{3.0} & \bf \normalsize{3.0} & \bf \normalsize{4.2} & \bf \normalsize{1.7} & \bf \normalsize{1.8} & \bf \normalsize{1.0} & \bf \normalsize{1.2} \\
\hline

 & $K \rightarrow \pi \nu \bar{\nu}$ & & $\tau \rightarrow Ke$ & $K \rightarrow \pi \nu \bar{\nu}$ & $K \rightarrow \pi \nu \bar{\nu}$ & $\tau \rightarrow Ke$ &  \\
   \small{1~2} & $5.8 \times 10^{-4}$ & & $6.3$ & $2.9 \times 10^{-4}$ & $2.9 \times 10^{-4}$ & $3.2$ &  \\
        & \bf \normalsize{4.0} & \bf \normalsize{4.0} \cellcolor{orange} & \bf
        \normalsize{5.0} \cellcolor{orange} & \bf \normalsize{2.1} & \bf
        \normalsize{3.5} & \bf \normalsize{2.3} \cellcolor{orange} & \bf
        \normalsize{3.1} \cellcolor{orange} \\
\hline

    & & & $B \rightarrow \tau \bar{e}$ & $V_{ub}$ & $B \rightarrow \tau \bar{e}$ & $B \rightarrow \tau \bar{e}$ &  \\
    \small{1~3} & {\bf \large{$\ast$}} & {\bf \large{$\ast$}} & $0.3$ & $0.12$ & $0.13$ & $0.13$ & {\bf \large{$\ast$}}\\
    & & & \bf \normalsize{5.3} & \bf \normalsize{2.7} & \bf
    \normalsize{4.2} & \bf \normalsize{4.2} & \\


\hline
\hline
& $K \rightarrow \pi \nu \bar{\nu}$ & & $\tau \rightarrow Ke$ & $K \rightarrow \pi \nu \bar{\nu}$ & $K \rightarrow \pi \nu \bar{\nu}$ & $\tau \rightarrow Ke$ &  \\
   \small{2~1} & $5.8 \times 10^{-4}$ & & $6.3$ & $2.9 \times 10^{-4}$ & $2.9 \times 10^{-4}$ & $3.2$ &  \\
        & \bf \normalsize{4.2} & \bf \normalsize{4.2} \cellcolor{orange} & \bf
        \normalsize{5.5} \cellcolor{orange} & \bf \normalsize{2.3} & \bf
        \normalsize{1.8} & \bf \normalsize{1.1} \cellcolor{orange} & \bf
        \normalsize{1.2} \cellcolor{orange} \\
\hline

 & $\tau \rightarrow 3e$ & $\tau \rightarrow 3e$ & $\tau \rightarrow 3e$ & $\tau \rightarrow 3e$ & $\tau \rightarrow 3e$ & $\tau \rightarrow 3e$ & $\tau \rightarrow 3e$  \\
   \small{2~2} & $5.0$  & $5.0$ & $17.0$ & $14.0$ & $9.0$ & $4.0$ & $3.0$ \\
        & \bf \normalsize{10.8} & \bf \normalsize{10.9} & \bf \normalsize{8.6}
        \cellcolor{orange} & \bf \normalsize{3.9} \cellcolor{orange} & \bf
        \normalsize{4.5} \cellcolor{orange} & \bf \normalsize{3.5}
        \cellcolor{orange} & \bf \normalsize{5.5} \\
\hline

    & & & $B \rightarrow \bar{\tau} eX$ & $B \rightarrow \bar{\tau} eX$ & $B \rightarrow \bar{\tau} eX$ & $B \rightarrow \bar{\tau} eX$ &  \\
     \small{2~3} & {\bf \large{$\ast$}} & {\bf \large{$\ast$}} & $14.0$ & $7.2$ & $7.2$ & $7.2$ & {\bf \large{$\ast$}}\\
      &  & & \bf \normalsize{9.3} \cellcolor{orange} & \bf \normalsize{4.7}
      \cellcolor{orange} &
      \bf\normalsize{6.3} \cellcolor{orange} & \bf \normalsize{6.3}
      \cellcolor{orange} &
      \\


\hline
\hline
 & & & $B \rightarrow \tau \bar{e}$ & $B \rightarrow \tau \bar{e}$ & $B \rightarrow \tau \bar{e}$ & $B \rightarrow \tau \bar{e}$ &    \\
 \small{3~1}& {\bf \large{$\ast$}} & {\bf \large{$\ast$}} & $0.3$ & $0.13$ & $0.13$ & $0.13$ &  {\bf \large{$\ast$}}\\
     
 & & & \bf \normalsize{6.3} & \bf \normalsize{3.1} & \bf \normalsize{1.9} & \bf \normalsize{1.9} & \\
\hline

 & & & $B \rightarrow \bar{\tau}eX$ & $B \rightarrow \bar{\tau}eX$ & $B \rightarrow \bar{\tau}eX$ & $B \rightarrow \bar{\tau}eX$ & \\
 \small{3~2} & {\bf \large{$\ast$}} & {\bf \large{$\ast$}} & $14.0$ & $7.2$ &
 $7.2$ & $7.2$ & {\bf \large{$\ast$}} \\
 & & & \bf \normalsize{12.6} \cellcolor{orange} & \bf \normalsize{6.4}
 \cellcolor{orange} &  \bf \normalsize{4.9} \cellcolor{orange} & \bf
 \normalsize{4.9} \cellcolor{orange} &   \\
\hline

 & & & $\tau \rightarrow 3e$ & $\tau \rightarrow 3e$ & $\tau \rightarrow 3e$ & $\tau \rightarrow 3e$ &    \\
 \small{3~3} & {\bf \large{$\ast$}} & {\bf \large{$\ast$}} & $17.0$ & $14.0$ & $9.0$ & $4.0$ & {\bf \large{$\ast$}}  \\
 & & & \bf \normalsize{15.2} \cellcolor{orange} & \bf \normalsize{7.8}
 \cellcolor{orange} & \bf \normalsize{8.1} \cellcolor{orange} & \bf \normalsize{8.1} &  \\ 
\hline


\end{tabular}
\end{tiny}
\end{center}
\caption[Limits on $\lambda_{eq_i}\lambda_{\tau
  q_j}/m_{\rm LQ}^2$ for high mass $F=2$ LQs]{
  Limits at 95\% CL on $\lambda_{eq_i}\lambda_{\tau
  q_j}/m_{\rm LQ}^2$ for $F=2$ leptoquarks (bold).
  Combinations of $i$ and $j$ shown in the first column denote 
  the quark generation coupling to the electron and tau respectively. 
  In each cell the first two rows show the process providing the most stringent limit 
  from low energy experiments. 
  The cases marked with '$\ast$' refer to scenarios involving a top
  quark, not considered in the present analysis. Highlighted H1 limits are more stringent than those from the
  corresponding low energy experiment.}
\label{F2tauHighMassLQ}
\end{table}

\end{document}